\documentclass[aps,prc,twocolumn,raggedbottom]{revtex4-1}%
\pdfoutput=1
\usepackage[utf8]{inputenc} 
\usepackage[T1]{fontenc}    
\usepackage{graphicx}
\newcommand{\inputpgf}[1]{\includegraphics{img/#1.pdf}}
\usepackage{amsmath, amssymb} 
\everymath{\displaystyle}     
\renewcommand{\Im}{\operatorname{Im}}     
\DeclareMathOperator{\bigo}{\mathcal{O}}  
\newcommand{\D}{\mathop{}\!\mathrm{d}}    
\newcommand{\E}{\mathop{}\!\mathrm{e}}    
\newcommand{\I}{\mathrm{i}}               
\newcommand{\anbohr}{a_{\mathrm{B}}}      
\newcommand{\Ryd}{\mathrm{Ry}}            
\newcommand{\der}[3][]{\frac{\D^{#1} #2}{\D #3^{#1}}}           
\newcommand{\braket}[2]{\left\langle#1\middle|#2\right\rangle}  
\newcommand{\abs}[1]{\left|#1\right|}                           
\newcommand\energy{\epsilon} 
\newcommand\fermiint{F}      
\newcommand\wmax{\bar{w}}    
\newcommand\hypm[3]{{}_1F_1\!\left(\substack{#1\\#2};#3\right)}
\newcommand\regm[3]{\underline{M}\!\left(\substack{#1\\#2};#3\right)}
\newcommand\hypu[3]{U\!\left(\substack{#1\\#2};#3\right)}
\newcommand\hypf[4]{{}_2F_1\!\left(\substack{#1,#2\\#3};#4\right)}
\newcommand\regf[4]{\underline{F}\!\left(\substack{#1,\,#2\\#3};#4\right)}
\newcommand\respole[1]{(-140 #1 467\,\I)~\mathrm{keV}}
\usepackage{hyperref}  
\hypersetup{           
	pdfborderstyle={/S/U/W 1}, 
	pdfinfo={%
		Author       = {David Gaspard, et al. (ULB)},
		Title        = {Complex-energy analysis of proton-proton fusion},
		Subject      = {Nuclear Physics and Nucleosynthesis},
		CreationDate = {D:20181009160000}, 
	}
}

\begin{document}
\title{Complex-energy analysis of proton-proton fusion}
\author{David Gaspard}\email[FNRS Aspirant,~]{dgaspard@ulb.ac.be}
\author{Jean-Marc Sparenberg}\email{jmspar@ulb.ac.be}
\author{Quentin Wenda}
\author{Daniel Baye}  
\affiliation{Nuclear Physics and Quantum Physics, CP229, Universit\'e libre de Bruxelles (ULB), \'Ecole polytechnique, B-1050 Brussels, Belgium}
\date{September 27, 2019}

\begin{abstract}
An analysis of the astrophysical $S$~factor of the proton-proton weak capture ($\mathrm{p}+\mathrm{p}\rightarrow {}^2\mathrm{H}+\mathrm{e}^++\nu_{\mathrm{e}}$) is performed on a large energy range covering solar-core and early Universe temperatures.
The measurement of~$S$ being physically unachievable, its value relies on the theoretical calculation of the matrix element~$\Lambda$.
Surprisingly, $\Lambda$ reaches a maximum near $0.13~\mathrm{MeV}$ that has been unexplained until now.
A model-independent parametrization of~$\Lambda$ valid up to about $5~\mathrm{MeV}$ is established on the basis of recent effective-range functions.
It provides an insight into the relationship between the maximum of~$\Lambda$ and the proton-proton resonance pole at~$\respole{-}$ from analytic continuation.
In addition, this parametrization leads to an accurate evaluation of the derivatives of~$\Lambda$, and hence of~$S$, in the limit of zero energy.
\end{abstract}
\maketitle

\section{Introduction}
The proton-proton fusion reaction ($\mathrm{p}+\mathrm{p}\rightarrow {}^2\mathrm{H}+\mathrm{e}^++\nu_{\mathrm{e}}$), also known as the proton-proton weak capture, is a fundamental process in nuclear astrophysics.
It is the starting point of the proton-proton chain for stellar nucleosynthesis in hydrogen-burning stars.
Its cross section $\sigma(E)$ is usually expressed in terms of the astrophysical factor $S(E)$ at the two-proton center-of-mass energy~$E$.
Unfortunately, this cross section is so small at typical astrophysical temperature ($E\lesssim 0.01~\mathrm{MeV}$), that a reliable measurement cannot be achieved with enough statistics, even above the Coulomb barrier (about $0.2~\mathrm{MeV}$).
A theoretical prediction of $S$ is therefore required.
\par The first calculation of $S$ at zero energy was proposed by Bethe and Critchfield~\cite{Bethe1938}.
They also introduced the dimensionless weak capture matrix element $\Lambda$ at zero angular momentum from which they deduced $S$.
Thereafter, the accuracy of $S(0)$ was improved by Salpeter~\cite{Salpeter1952} and Bahcall and his coworkers~\cite{Bahcall1969} using effective-range theory.
In the 1990s, several authors calculated $S(0)$ from the nucleon-nucleon wave functions computed in potential models~\cite{Bahcall1992, Kamionkowski1994}.
More recently, systematic computations of $S(0)$ were performed in pionless effective field theory from next-to-leading order (NLO) of the momentum expansion~\cite{Kong2001, Ando2008} up to N\textsuperscript{4}LO~\cite{Butler2001b}.
In parallel, efforts were made in chiral effective field theory to reduce the uncertainty on $S(0)$ by adding two-body corrections to the Gamow-Teller operator adjusted with data for tritium $\beta$-decay~\cite{Carlson1991, Park2003, Schiavilla1998, Marcucci2013, *Marcucci2014a}.
As the uncertainty on $S(0)$ has diminished since early works, the small contribution of its energy derivatives $S'(0)$ and $S''(0)$ has become important for nuclear astrophysics~\cite{Adelberger2011, Marcucci2013, *Marcucci2014a, ChenJW2013, Acharya2016}.
In this regard, Adelberger~\emph{et al.} recommended a calculation of $S''(0)$ to be undertaken~\cite{Adelberger2011}.
\par The calculation of these derivatives raises the question of whether $S$ is analytic in the neighborhood of zero energy.
Such an analysis is still missing in the literature.
Yet, some peculiarities of the proton-proton scattering are known.
In~1980, Kok highlighted the presence of a sub-threshold resonance pole at about~$\respole{-}$ in the ${}^1S_0$ channel~\cite{Kok1980}.
Up to a complex phase, this pole lies in an energy range corresponding to early Universe temperatures below the nucleosynthesis freeze-out point ($E\lesssim 1~\mathrm{MeV}$)~\cite{Tanabashi2018}.
Therefore, the derivatives of $S$ are likely to be influenced by the relative closeness of this pole.
In addition, it is known that the Coulomb interaction between two protons generates a sequence of poles at complex energy which accumulate to the zero-energy point~\cite{Kok1980, GaspardD2018a}.
These poles may affect the polynomial extrapolation used in Refs.~\cite{Schiavilla1998, Marcucci2013, *Marcucci2014a, Acharya2016} to determine $S(0)$, $S'(0)$, and $S''(0)$ from the numerical computation of $S(E)$ on an energy range that does not include $E=0$.
In this context, the issue is to understand the influence of all these structures on the astrophysical $S$ factor.
\par The purpose of this work is to obtain an efficient model-independent parametrization of $S$, valid on a large energy range ($E\lesssim5~\mathrm{MeV}$), and able to impose constraints on its series expansion at $E=0$.
Our parametrization must also describe the resonance pole at~$\respole{-}$ and the Coulomb poles.
To do so, we resort to a recently introduced effective-range function (ERF), namely the $\Delta$ function~\cite{RamirezSuarez2017, GaspardD2018a, Baye2000a}, which has only been considered useful for heavier systems until now~\cite{Blokhintsev2017, *Blokhintsev2018a, *Blokhintsev2018b}.
This approach is motivated by the efficiency of the ERFs at describing the energy-dependent shape of the proton-proton wave function up to a few MeVs.
It is aimed to reach a much better accuracy on the values of $S(0)$, $S'(0)$, and $S''(0)$ than with the polynomial extrapolation performed in Refs.~\cite{Schiavilla1998, Marcucci2013, *Marcucci2014a, Acharya2016}.
Finally, our results will be verified with the nucleon-nucleon wave functions computed in different local potential models, namely Av18~\cite{Wiringa1995}, Reid93~\cite{Stoks1994}, and NijmII~\cite{Stoks1994}.
In particular, these potentials are intended to validate the model independence of our parametrization.
\par This paper is organized as follows.
Section~\ref{sec:pp-weak-capture} presents the parametrization of the astrophysical $S$ factor, first from analytical approximations of the nucleon-nucleon wave functions in Subsec.~\ref{subsec:wavefunctions}, and then in a model-independent way in Subsec.~\ref{subsec:parametrization}.
The analysis of $S$ at complex energies is discussed in Sec.~\ref{sec:complex-analysis} and supplemented by graphical illustrations.
The numerical values of the logarithmic derivatives of $S$ are shown in Sec.~\ref{sec:derivatives}.
Section~\ref{sec:conclusion} is devoted to a conclusion.
Detailed calculations of the Fermi phase-space integral and the Coulomb integrals are given in Appendices~\ref{app:fermi-integral} and~\ref{app:coulomb-integrals}, respectively.

\section{Proton-proton weak capture\label{sec:pp-weak-capture}}
\subsection{Astrophysical $S$ factor\label{subsec:astro-s-factor}}
Let us start by defining the astrophysical $S$ factor studied in this paper.
After integrating out the emitted leptons, the proton-proton weak capture cross section is known to be given to a good approximation by~\cite{Salpeter1952, Bahcall1969, Baye2013}
\begin{equation}\label{eq:pp-cross-section-1}
\sigma(E) = \frac{3m_{\rm e}c^2(\lambda g)^2}{\pi^2Ek}\fermiint(E+Q)\abs{\int_0^{\infty}u_{\rm d}(r)\,u_{\rm pp}(E,r)\D r}^2  \:,
\end{equation}
where $k=\sqrt{m_{\rm p}E}/\hbar$ is the proton-proton wave number, and $u_{\rm d}$ and $u_{\rm pp}$ are the radial $S$-wave components of the deuteron and proton-proton wave functions respectively.
The contribution from higher-order partial waves can be neglected in the low-energy approximation~\cite{Acharya2019}.
The parameter $\lambda=1.2724(23)$ is the weak axial/vector ratio, $g=G_{\rm F}\abs{V_{\rm ud}}(m_{\rm e}c^2)^2/(\hbar c)^3=2.96707(64)\times 10^{-12}$ is the dimensionless weak coupling constant for neutrons, $G_{\rm F}$ is the Fermi constant of muon decay, and $V_{\rm ud}$ is the first element of the CKM quark mixing matrix. All the fundamental constants are taken from Ref.~\cite{Tanabashi2018}.
The quantity $\fermiint(E+Q)$ in Eq.~\eqref{eq:pp-cross-section-1} is known as the Fermi phase-space integral, which accounts for the electric repulsion of the emitted positron in a relativistic framework. It is discussed in more detail in Appendix~\ref{app:fermi-integral}.
\par It turns out that the overlap integral between $u_{\rm d}$ and $u_{\rm pp}$ in Eq.~\eqref{eq:pp-cross-section-1} vanishes in the zero energy limit ($E\rightarrow 0$).
This vanishing behavior originates from the cancellation of the proton-proton radial wave function $u_{\rm pp}$ at $E=0$ due to the Coulomb barrier. 
This behavior can be factored out of the overlap integral in defining the dimensionless orbital matrix element $\Lambda(E)$ as~\cite{Salpeter1952, Bahcall1969}
\begin{equation}\label{eq:weak-capture-lambda}
\Lambda(E) = \frac{\anbohr b^{3/2}}{\sqrt{8}} \frac{2\eta}{C_{\eta,0}}\int_0^\infty u_{{\rm d},0}(r)\,u_{{\rm pp},0}(E,r) \D r  \:,
\end{equation}
where $C_{\eta,0}=\sqrt{2\pi\eta/(\E^{2\pi\eta}-1)}$ is the Coulomb normalization coefficient, $\eta=1/(\anbohr k)$ is the Sommerfeld parameter, and $\anbohr=\hbar c/(\alpha m_{\rm p}c^2/2)=57.6398~\mathrm{fm}$ is the proton-proton Bohr radius.
The other constants are the deuteron binding wave number $b=\sqrt{2m_{\rm pn}B_{\rm d}}/\hbar=0.231606~\mathrm{fm}^{-1}$~\cite{Bahcall1969, Adelberger2011}, the proton-neutron reduced mass $m_{\rm pn}=m_{\rm p}m_{\rm n}/(m_{\rm p}+m_{\rm n})$, and the deuteron binding energy $B_{\rm d}=2.22457~\mathrm{MeV}$.
In contrast to the overlap integral, $\Lambda(E)$ has a finite limit at $E=0$.
With these notations, the cross section~\eqref{eq:pp-cross-section-1} becomes after some simplifications
\begin{equation}\label{eq:pp-cross-section-2}
\sigma(E) = \frac{1}{(\E^{2\pi\eta}-1)\,E} \frac{12m_{\rm e}c^2(\lambda g)^2}{\pi\anbohr b^3}\fermiint(E+Q) \abs{\Lambda(E)}^2  \:.
\end{equation}
This expression~\eqref{eq:pp-cross-section-2} suggests the most natural definition of the astrophysical $S$ factor, that is
\begin{equation}\label{eq:astro-def}
S(E) = (\E^{2\pi\eta}-1)\,E\,\sigma(E)  \:.
\end{equation}
This definition~\eqref{eq:astro-def} of $S$ is assumed in this work.
Strictly speaking, the definition~\eqref{eq:astro-def} does not reduce to the original definition due to Salpeter~\cite{Salpeter1952}
\begin{equation}\label{eq:astro-std}
S_{\rm std}(E) = \E^{2\pi\eta}\,E\,\sigma(E)  \:,
\end{equation}
and nowadays considered as standard in stellar astrophysics.
We point out that Eq.~\eqref{eq:astro-def} is considered in Sec.~II of Bahcall's and May's paper~\cite{Bahcall1969}, even though they actually defined $S$ by Eq.~\eqref{eq:astro-std}.
\par Anyway, at sufficiently low energy, the definitions~\eqref{eq:astro-def} and~\eqref{eq:astro-std} coincide.
We notice indeed that the approximation $\E^{2\pi\eta}-1\simeq\E^{2\pi\eta}$ is valid within less than $1\%$ for
\begin{equation}
E\leq \left(\frac{2\pi}{\ln(100)}\right)^2\Ryd \simeq 1.86\:\Ryd  \:,
\end{equation}
where $\Ryd$ is the nuclear Rydberg energy which is equal to $\alpha^2m_{\rm p}c^2/4=12.4911~\mathrm{keV}$ in the proton-proton system.
Since we consider energies much higher than the Rydberg energy in this work, definition~\eqref{eq:astro-def} is preferred.
Thus, we understand Eq.~\eqref{eq:astro-std} as the low-energy approximation of Eq.~\eqref{eq:astro-def}.
\par Furthermore, it should be noted that definition~\eqref{eq:astro-def} does not affect any of the values $S(0)$, $S'(0)$, and $S''(0)$ presented in the main text compared to $S_{\rm std}$.
Indeed, the relative error between the two definitions displays an essential singularity at $E=0$ which cancels all its derivatives in the limit $E\xrightarrow{>}0$.
Therefore, the derivatives of $S$ at $E=0$ obtained in this work are necessarily equal to the derivatives of $S_{\rm std}$ at $E=0$.
\par Finally, using Eqs.~\eqref{eq:pp-cross-section-2} and~\eqref{eq:astro-def}, one finds the expression
\begin{equation}\label{eq:astrophysical-factor}
S(E) = \frac{12m_{\rm e}c^2(\lambda g)^2}{\pi\anbohr b^3}\fermiint(E+Q)\abs{\Lambda(E)}^2  \:.
\end{equation}
The currently recommended value of $S(0)$ is $4.01(4)\times 10^{-23}~\mathrm{MeV}\,\mathrm{fm}^2$~\cite{Adelberger2011}.

\subsection{Wave function-based parametrization\label{subsec:wavefunctions}}
Most of the energy dependence in the overlap integral~\eqref{eq:weak-capture-lambda} comes from the proton-proton wave function $u_{\rm pp}$.
This function is normalized such that it tends to a sine wave of unit amplitude for $r\rightarrow\infty$.
Its asymptotic behavior reads
\begin{equation}\label{eq:diproton-far-field}
u_{{\rm pp},0}(r) \xrightarrow{r\rightarrow\infty} F_{\eta,0}\cos\delta_0 + G_{\eta,0}\sin\delta_0  \:,
\end{equation}
where $F_{\eta,\ell}(kr)$ and $G_{\eta,\ell}(kr)$ are the standard Coulomb wave functions~\cite{Olver2010}, and $\delta_0$ is the proton-proton $^1S_0$ phase shift.
Similarly, the asymptotic behavior of the deuteron $^3S_1$ wave function is
\begin{equation}\label{eq:deuteron-far-field}
u_{{\rm d},0}(r) \xrightarrow{r\rightarrow\infty} A\E^{-br}  \:.
\end{equation}
The normalization coefficient $A$ is found to be $0.8850(5)~\mathrm{fm}^{-1/2}$ using three different potential models (Av18, Reid93, and NijmII)~\cite{Wiringa1995, Stoks1994}.
It turns out that the asymptotic regime in Eqs.~\eqref{eq:diproton-far-field} and~\eqref{eq:deuteron-far-field} is already reached for $r\gtrsim 2~\mathrm{fm}$.
Since the spatial extent of the deuteron is much larger than $2~\mathrm{fm}$, it is relevant~\cite{Salpeter1952, Bahcall1969} to calculate~\eqref{eq:weak-capture-lambda} from these asymptotic behaviors.
\par Inserting the asymptotic behaviors~\eqref{eq:diproton-far-field} and~\eqref{eq:deuteron-far-field} into Eq.~\eqref{eq:weak-capture-lambda} leads to the Laplace transforms of the Coulomb functions, which are known analytically in terms of Gauss hypergeometric functions.
However, this calculation does not consider the short-range behavior of the nucleon-nucleon wave functions due to the nuclear potential.
\par One efficient way of including the short-range contribution in $\Lambda$ is to use analytical approximations of the nucleon wave functions.
This method was used by many authors, especially to approximate the deuteron wave function on the basis of a series of exponential functions.
Such approximations are known as Hulthén-type wave functions~\cite{Moravcsik1958, Kottler1964, McGee1966, Humberston1970, Oteo1988}.
\par
\begin{figure}[ht]%
\inputpgf{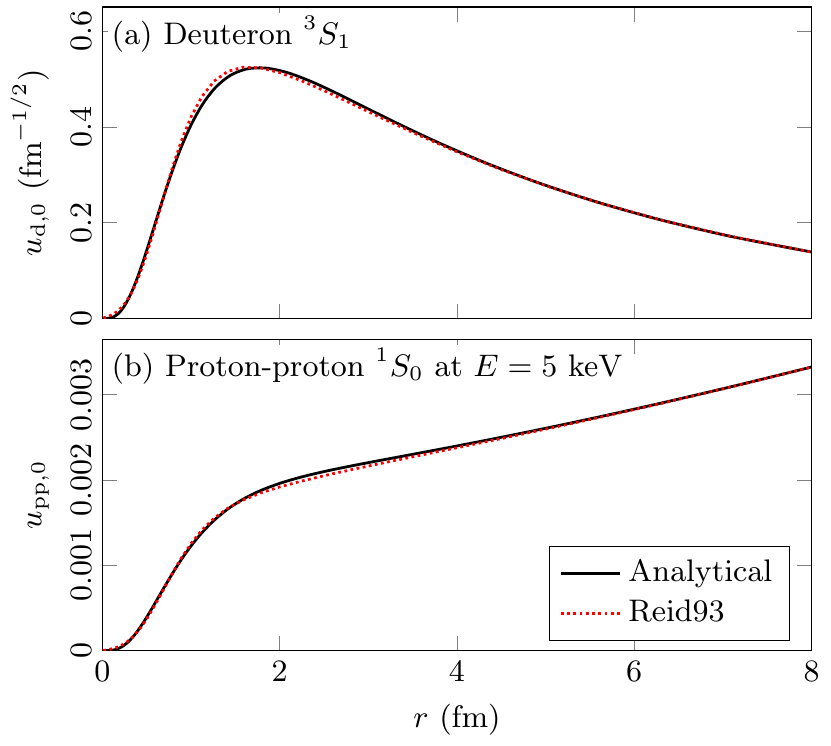}
\caption{Comparison between the analytical approximations of the nucleon-nucleon wave functions and the wave functions computed with the Reid93 potential~\cite{Stoks1994}.}
\label{fig:plot-nucleons}
\end{figure}

 We propose to use the following approximation of the $^3S_1$ bound state of the deuteron:
\begin{equation}\label{eq:exp-fit-deuteron}
u_{{\rm d},0}(r) = A(1-\E^{-s_{\rm d}br})^{\nu_{\rm d}}\E^{-br}  \:.
\end{equation}
The shape parameters $s_{\rm d}$ and $\nu_{\rm d}$ are fitted to the deuteron wave function.
It is worth noting that $A$ is related to the shape parameters $s_{\rm d}$ and $\nu_{\rm d}$ in Eq.~\eqref{eq:exp-fit-deuteron} by the normalization condition
\begin{equation}\label{eq:exp-fit-amplitude}
A = \sqrt{N_{\rm S}\frac{2b\,\Gamma(2\nu_{\rm d}+2s_{\rm d}^{-1}+1)}{\Gamma(2\nu_{\rm d}+1)\,\Gamma(2s_{\rm d}^{-1}+1)}}  \:,
\end{equation}
where $N_{\rm S}$ is the ${}^3S_1$-state probability given by $\braket{u_{{\rm d},0}}{u_{{\rm d},0}}=94.30(6)\%$ for the three potential models (Av18, Reid93, NijmII).
The fitted values of the shape parameters subject to the constraint~\eqref{eq:exp-fit-amplitude} are shown in Table~\ref{tab:fit-nucleons}.
As one can see in Fig.~\ref{fig:plot-nucleons}(a), the approximation~\eqref{eq:exp-fit-deuteron} of the deuteron wave function is remarkably accurate.
The root-mean-square deviation from the Reid93 ${}^3S_1$ wave function is about $0.015$.
This accuracy is good enough for our needs.%
\begin{table}[ht]%
\begin{tabular}{l*{4}{c}}\hline
                         & $s_{\rm d}$ & $\nu_{\rm d}$ & $s_{\rm p}$ & $\nu_{\rm p}$  \\ \hline
Av18~\cite{Wiringa1995}  & $8.43$      & $3.61$        & $9.81$      & $4.22$         \\
Reid93~\cite{Stoks1994}  & $8.45$      & $3.63$        & $9.55$      & $3.92$         \\
NijmII~\cite{Stoks1994}  & $8.78$      & $4.00$        & $9.59$      & $3.95$         \\ \hline
\end{tabular}%
\caption{Shape parameters in different potential models for the S-state two-nucleons wave functions.}\label{tab:fit-nucleons}
\end{table}%
\par The same kind of parametrization can be applied to the proton-proton wave function:
\begin{equation}\label{eq:exp-fit-diproton}
u_{{\rm pp},0}(r) = (1-\E^{-s_{\rm p}br})^{\nu_{\rm p}} (F_{\eta,0}\cos\delta_0 + G_{\eta,0}\sin\delta_0)  \:.
\end{equation}
In contrast to the deuteron wave function, this wave function depends on the energy of the incoming protons.
Therefore, the shape parameters $s_{\rm p}$ and $\nu_{\rm p}$ are likely to vary with the energy.
However, we will neglect these variations on the considered energy range because of the great depth of the nuclear potentials.
The fitted values of the shape parameters are shown in Table~\ref{tab:fit-nucleons} for different potential models.
\par The advantage of the expressions~\eqref{eq:exp-fit-deuteron} and~\eqref{eq:exp-fit-diproton} is the reduction to the known Laplace transforms of the Coulomb functions, if the shape factors $(1-\E^{-s_{\rm d}x})^{\nu_{\rm d}}$ and $(1-\E^{-s_{\rm p}x})^{\nu_{\rm p}}$ are expanded in binomial series.
The product function then becomes
\begin{equation}\label{eq:product-binomial-expansion}
u_{{\rm d},0}\,u_{{\rm pp},0} = \sum_{i,j}^\infty c_{i,j} \E^{-\beta_{i,j}br}(F_{\eta,0}\cos\delta_0 + G_{\eta,0}\sin\delta_0)  \:.
\end{equation}
The integer indices $i$ and $j$ in the expansion~\eqref{eq:product-binomial-expansion} run over the terms of the binomial expansion of $(1-\E^{-s_{\rm d}x})^{\nu_{\rm d}}$ and $(1-\E^{-s_{\rm p}x})^{\nu_{\rm p}}$ respectively.
The variable $\beta_{i,j}$ takes the values
\begin{equation}
\beta_{i,j} = 1 + is_{\rm d} + js_{\rm p}  \:,
\end{equation}
and the corresponding coefficient is
\begin{equation}\label{eq:hulthen-coeff}
c_{i,j} = (-1)^{i+j} \binom{\nu_{\rm d}}{i} \binom{\nu_{\rm p}}{j}  \:.
\end{equation}
\par
\begin{figure}[ht]%
\inputpgf{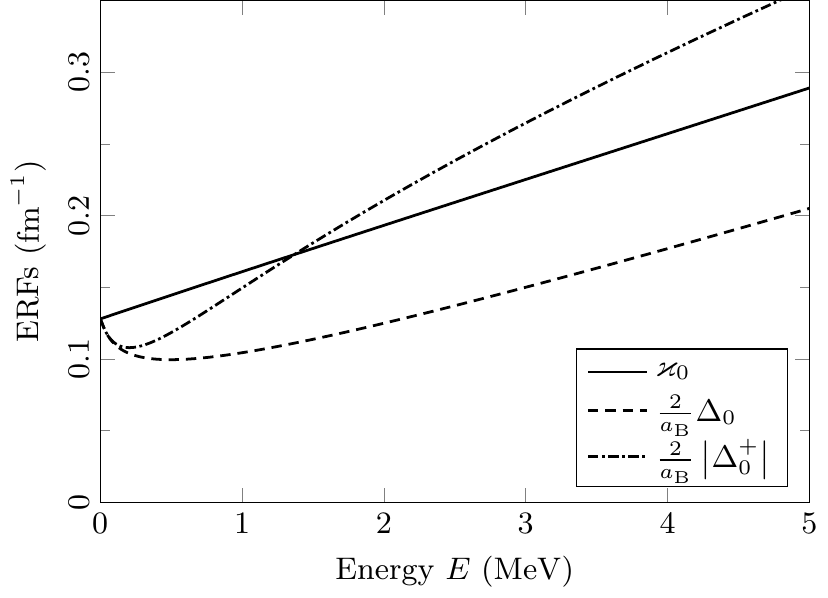}
\caption{Effective-range functions of the proton-proton ${}^1S_0$ scattering for the Reid93 potential~\cite{Stoks1994}.
The other potential models provide very close curves at this scale.}
\label{fig:plot-erf}
\end{figure}

 One important issue with the overlap integral~\eqref{eq:weak-capture-lambda} is the strongly vanishing behavior of the proton-proton wave function when the energy decreases.
The wave function $u_{{\rm pp},0}(r)$, normalized according to Eq.~\eqref{eq:diproton-far-field}, tends to zero as $\eta^{-1/2}\E^{-\pi\eta}$~\cite{Baye2013}.
This behavior is due to the asymptotic normalization of the Coulomb wave functions $F_{\eta,\ell}$ and $G_{\eta,\ell}$ that are constrained to sine wave of unit amplitude.
In Eq.~\eqref{eq:weak-capture-lambda}, this cancellation is compensated by the prefactor $2\eta/C_{\eta,0}$.
In order to factorize the cancellation of $u_{{\rm pp},0}(r)$ out of the overlap integral, we rewrite the Coulomb wave functions $F_{\eta,\ell}$ and $G_{\eta,\ell}$ of Eq.~\eqref{eq:diproton-far-field} in terms of the modified Coulomb functions $\Phi_{\eta,\ell}$ and $\Psi_{\eta,\ell}$ introduced in Refs.~\cite{GaspardD2018a, GaspardD2018b}.
Contrary to the standard Coulomb functions, these functions have the advantage of being analytic in the complex plane of the energy.
In particular, they tend toward nonzero functions of $r/\anbohr$ at zero energy.
The function $\Phi_{\eta,\ell}$ is related to $F_{\eta,\ell}$ by~\cite{GaspardD2018b}
\begin{equation}\label{eq:coulomb-F-from-Phi}
F_{\eta,\ell}(kr) = \frac{C_{\eta,\ell}\Gamma(2\ell+2)}{(2\eta)^{\ell+1}}\Phi_{\eta,\ell}(kr)  \:,
\end{equation}
where $C_{\eta,\ell}$ is the general Coulomb normalization coefficient that reads~\cite{Olver2010, GaspardD2018b}
\begin{equation}\label{eq:coulomb-c}
C_{\eta,\ell} = \frac{(2\eta)^\ell}{\Gamma(2\ell+2)}\sqrt{\frac{2\pi\eta\,w_{\eta,\ell}}{\E^{2\pi\eta} - 1}} \:.
\end{equation}
In Eq.~\eqref{eq:coulomb-c}, the function $w_{\eta,\ell}$ is defined by
\begin{equation}
w_{\eta,\ell} = \prod_{j=0}^\ell\left(1+\frac{j^2}{\eta^2}\right)  \quad\text{and}~w_{\eta,0}=1  \:.
\end{equation}
The function $G_{\eta,\ell}$ can be expressed in terms of $\Phi_{\eta,\ell}$ and $\Psi_{\eta,\ell}$ as~\cite{GaspardD2018b}
\begin{equation}\label{eq:coulomb-G-from-Psi}
G_{\eta,\ell} = \frac{C_{\eta,\ell}\Gamma(2\ell+2)}{(2\eta)^{\ell+1}} \frac{\E^{2\pi\eta}-1}{\pi}\!\left(\frac{\Psi_{\eta,\ell}}{w_{\eta,\ell}} + g_{\eta,\ell}\Phi_{\eta,\ell}\right)  \:,
\end{equation}
where $g_{\eta,\ell}$ is the Bethe function given by
\begin{equation}\label{eq:bethe-g}
g_{\eta,\ell} = \frac{\psi(\ell+1+\I\eta)+\psi(\ell+1-\I\eta)}{2} - \ln\eta  \:,
\end{equation}
and $\psi(z)=\Gamma'(z)/\Gamma(z)$ is the digamma function~\cite{Bethe1949, GaspardD2018a}.
Combining Eqs.~\eqref{eq:coulomb-F-from-Phi} and~\eqref{eq:coulomb-G-from-Psi}, the asymptotic behavior of the proton-proton wave function~\eqref{eq:diproton-far-field} becomes
\begin{equation}\label{eq:coulomb-eff-range-asym}\begin{split}
 & F_{\eta,0}\cos\delta_0 + G_{\eta,0}\sin\delta_0  \\
 & = \frac{C_{\eta,0}/(2\eta)}{\abs{\Delta^+_0(E)}}\!\left(\frac{\anbohr}{2}\varkappa_0\Phi_{\eta,0} + \Psi_{\eta,0}\right)   \:.
\end{split}\end{equation}
In contrast to $F_{\eta,\ell}$ and $G_{\eta,\ell}$, the modified Coulomb functions $\Phi_{\eta,\ell}$ and $\Psi_{\eta,\ell}$ are free of singularity at zero energy.
This property is crucial in the analysis of $\Lambda$ at complex energy, especially near $E=0$.
In Eq.~\eqref{eq:coulomb-eff-range-asym}, $\varkappa_0$ is the standard ERF of the proton-proton ${}^1S_0$ scattering.
Its first-order expansion in $E$ provides an accurate parametrization of the phase shift $\delta_0$ over a large energy range ($E\lesssim 5~\mathrm{MeV}$)
\begin{equation}\label{eq:eff-range-function-0}
\varkappa_0(E) = \frac{2}{\anbohr}\!\left(\frac{\pi\cot\delta_0}{\E^{2\pi\eta}-1} +g_{\eta,0}\right) \simeq \frac{-1}{\alpha_0} + \frac{r_0}{2}\frac{m_{\rm p}}{\hbar^2}E  \:.
\end{equation}
The parameters $\alpha_0$ and $r_0$ are respectively the scattering length and the effective range.
The modulus of the modified ERF $\Delta^+_0$~\cite{GaspardD2018a, Hamilton1973} also appears in Eq.~\eqref{eq:coulomb-eff-range-asym}.
This phase-shift-dependent function is related to $\varkappa_0$ by
\begin{equation}\label{eq:diproton-delta-plus-0}
\Delta^+_0(E) = \left(\frac{\anbohr}{2}\varkappa_0-g_{\eta,0}\right) - \frac{\I\pi}{\E^{2\pi\eta}-1}  \:. 
\end{equation}
However, in contrast to $\varkappa_0$, the function $\Delta^+_0$ is singular at $E=0$ mostly because of $g_{\eta,0}$.
The bracket in Eq.~\eqref{eq:diproton-delta-plus-0} is also called the $\Delta_0$ function in Ref.~\cite{GaspardD2018a}.
In practical computation, $\Delta^+_0$ can be evaluated from the knowledge of the effective-range function $\varkappa_0$.
The square modulus of $\Delta^+_0$ can be expressed as
\begin{equation}\label{eq:diproton-delta-plus-2}
\abs{\Delta^+_0(E)}^2 = \left(\frac{\anbohr}{2}\varkappa_0 - g_{\eta,0}\right)^2 + \left(\frac{\pi}{\E^{2\pi\eta}-1}\right)^2  \:,
\end{equation}
as it will also enter the parametrization of $\Lambda$.
The expression~\eqref{eq:diproton-delta-plus-2} is the analytic continuation of $\abs{\Delta^+_0}^2$ to the complex plane of the energy.
Finally, all these ERFs are depicted in Fig.~\ref{fig:plot-erf}.
\par Using the expansion~\eqref{eq:product-binomial-expansion} of the product function, the overlap integral~\eqref{eq:weak-capture-lambda} splits into a series of Laplace transforms of the modified Coulomb functions $\Phi_{\eta,0}$ and $\Psi_{\eta,0}$.
These Laplace transforms are defined as follows
\begin{equation}\label{eq:laplace-coulomb}\begin{aligned}
\phi_{\beta,0}(\energy) & = \int_0^\infty \E^{-\beta x}\Phi_{\eta,0}(x\sqrt{\energy}) \D x  \:,\\
\psi_{\beta,0}(\energy) & = \int_0^\infty \E^{-\beta x}\Psi_{\eta,0}(x\sqrt{\energy}) \D x  \:,
\end{aligned}\end{equation}
where $x=br$ is the dimensionless radial coordinate, $\energy=k^2/b^2=E/B$ is the dimensionless proton energy, and $B=2m_{\rm pn}B_{\rm d}/m_{\rm p}$ is the deuteron binding energy corrected for the neutron-proton mass difference.
As shown in Appendix~\ref{app:coulomb-integrals}, the regular Coulomb integral is exactly given by
\begin{equation}\label{eq:laplace-coulomb-phi-0}
\phi_{\beta,0}(\energy) = \frac{\chi}{\beta^2+\energy}\exp\!\big(\chi\tfrac{\arctan(\sqrt{\energy}/\beta)}{\sqrt{\energy}}\big)  \:,
\end{equation}
where the dimensionless constant $\chi=2/(\anbohr b)=0.149816$ is due to Bahcall and May~\cite{Bahcall1969}.
The Laplace transform of $\Psi_{\eta,0}$ cannot be obtained in a simple form.
However, it is explained in Appendix~\ref{app:coulomb-integrals} that, as far as $\chi$ is small with respect to $1$, the following approximation holds for $E\lesssim 5~\mathrm{MeV}$
\begin{equation}\label{eq:laplace-coulomb-psi-0}
\psi_{\beta,0}(\energy) \simeq \Gamma(-1,\tfrac{\chi}{\beta})\,\phi_{\beta,0}(\energy)  \:,
\end{equation}
where $\Gamma(a,z)$ denotes the upper incomplete gamma function~\cite{Olver2010}.
\par Therefore, combining Eqs.~\eqref{eq:product-binomial-expansion} and~\eqref{eq:coulomb-eff-range-asym} into Eq.~\eqref{eq:weak-capture-lambda}, we get the expansion
\begin{equation}\label{eq:overlap-integral-split}
\tilde{\Lambda} = \frac{A/\sqrt{2b}}{\chi\abs{\Delta^+_0}} \sum_{i,j}^\infty c_{i,j}\left(\frac{\anbohr}{2}\varkappa_0\phi_{\beta_{i,j},0} + \psi_{\beta_{i,j},0}\right)  \:.
\end{equation}
The tilde over $\Lambda$ means that the result~\eqref{eq:overlap-integral-split} assumes the analytical approximations~\eqref{eq:exp-fit-deuteron} and~\eqref{eq:exp-fit-diproton}.
\par It is useful to separate the first term of the expansion~\eqref{eq:overlap-integral-split}, for which $\beta_{0,0}=1$, because it corresponds to the contribution of the far-field part of the nucleon wave functions.
One thus expects this contribution to be larger than the short-range part of the wave functions~\cite{Bahcall1969}.
We find convenient to introduce an energy-dependent function, denoted $\tilde{C}$, which gathers all the contributions of the short-range part of the wave functions.
The expansion~\eqref{eq:overlap-integral-split} is thus written as
\begin{equation}\label{eq:lambda-param-c}
\tilde{\Lambda} = \frac{A/\sqrt{2b}}{\chi\abs{\Delta^+_0}} \left[\left(\frac{\anbohr}{2}\varkappa_0 + \gamma\right)\phi_{1,0}(\energy) - \tilde{C}(\energy)\right]  \:,
\end{equation}
where $\gamma=\Gamma(-1,\chi)=4.28065$ is independent of the energy, and
\begin{equation}\label{eq:overlap-c-from-exp}
\tilde{C}(\energy) = -\sum_{i,j\neq0,0}^\infty c_{i,j}\left(\frac{\anbohr}{2}\varkappa_0\phi_{\beta_{i,j},0} + \psi_{\beta_{i,j},0}\right)  \:.
\end{equation}
The minus sign in front of the series of Eq.~\eqref{eq:overlap-c-from-exp} makes $\tilde{C}$ a positive function.
\begin{figure}[ht]%
\inputpgf{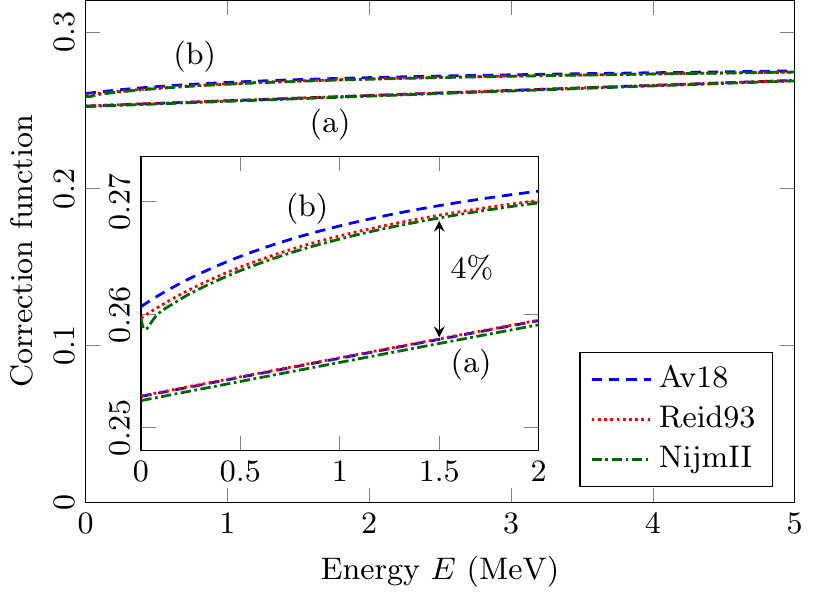}
\caption{Correction function for different potential models. 
The curves (a) depict the linear behavior of $\tilde{C}$ in Eq.~\eqref{eq:overlap-c-from-exp} using the data of Table~\ref{tab:fit-nucleons}.
The curves (b) show the numerical computation of $C$ from Eq.~\eqref{eq:c-from-lambda}.
The inset is an enlargement of the same curves.}
\label{fig:plot-overlap-correction}
\end{figure}%
The shape parameters of the nucleon wave functions in Table~\ref{tab:fit-nucleons} provide an estimation of $\tilde{C}$.
At zero energy, we get from Eq.~\eqref{eq:overlap-c-from-exp} the estimate $\tilde{C}(0)\simeq 0.253$.
The function $\tilde{C}$ built on Eq.~\eqref{eq:overlap-c-from-exp} is shown as curve (a) in Fig.~\ref{fig:plot-overlap-correction}.

\subsection{Model-independent parametrization\label{subsec:parametrization}}
Instead of relying on analytical approximations of the nucleon-nucleon wave functions, we can establish a general model-independent parametrization of $\Lambda$ inspired by Eq.~\eqref{eq:lambda-param-c} 
\begin{equation}\label{eq:lambda-param-c-gen}
\Lambda = \frac{A/\sqrt{2b}}{\chi\abs{\Delta^+_0}} \left[\left(\frac{\anbohr}{2}\varkappa_0 + \gamma\right)\phi_{1,0}(\energy) - C(\energy)\right]  \:.
\end{equation}
Indeed, this expression comes from the separation of the far-field part of the nucleon-nucleon wave functions from its short-range part $C$, and this far-field part is not supposed to change from a potential model to another.
Consequently, the parametrization~\eqref{eq:lambda-param-c-gen} is model independent, as far as the function $C$ can be fitted.
The function $C$ is defined by inverting Eq.~\eqref{eq:lambda-param-c-gen}
\begin{equation}\label{eq:c-from-lambda}
C(\energy) = \left(\frac{\anbohr}{2}\varkappa_0 + \gamma\right)\phi_{1,0}(\energy) - \frac{\chi\abs{\Delta^+_0}}{A/\sqrt{2b}}\Lambda(E)  \:,
\end{equation}
but computing $\Lambda$ with Eq.~\eqref{eq:weak-capture-lambda} on the basis of numerical wave functions.
At zero energy, we find $C(0)=0.260(1)$; the error being due to the uncertainty on potential models.
\par
\begin{figure}[ht]%
\inputpgf{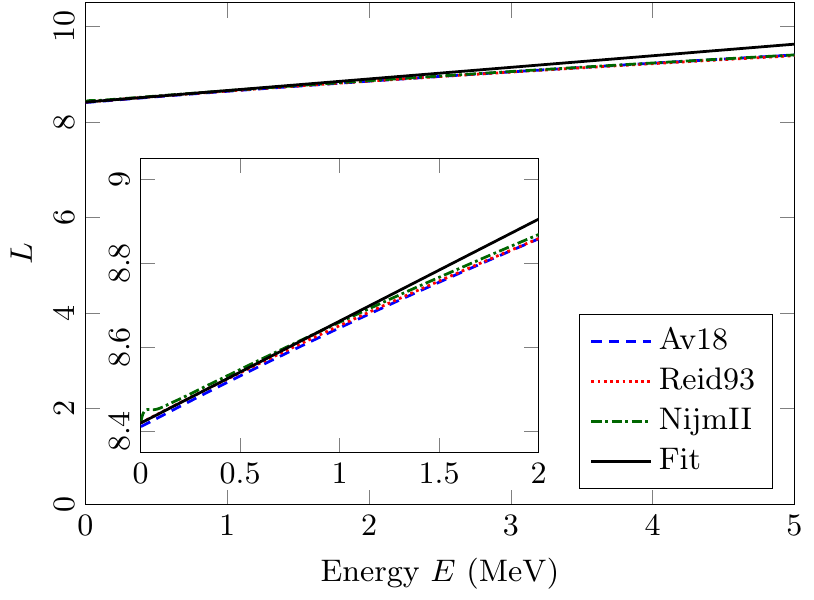}
\caption{Linearized matrix element $L$ of the weak capture. The vacuum polarization term of Av18 has been omitted because it leads to a spurious behavior at low energy. The linear fit has been achieved on $[300,600]~\mathrm{keV}$.}
\label{fig:plot-overlap-linearized}
\end{figure}

 The function $C$ in Eq.~\eqref{eq:c-from-lambda} is shown as curve (b) in Fig.~\ref{fig:plot-overlap-correction}.
Although they are close within less than $4\%$ on the considered energy range, curves (a) and (b) display different shapes.
We note that curve (b) deviates from a straight line, in contrast to curve (a).
This difference is due to the limitation of the assumption in Eq.~\eqref{eq:exp-fit-diproton} that the short-range part of the proton-proton wave function, i.e., the shaping factor $(1-\E^{-s_{\rm p}br})^{\nu_{\rm p}}$, does not depend on the energy.
Therefore, the use of function $C$ is limited to relatively low energy ($\lesssim 100~\mathrm{keV}$) where it can be considered linear.
In order to model the behavior over a large energy range ($\lesssim 5~\mathrm{MeV}$), we establish from Eqs.~\eqref{eq:laplace-coulomb-phi-0} and~\eqref{eq:lambda-param-c-gen} a parametrization of $\Lambda$ that is more convenient to practical applications
\begin{equation}\label{eq:lambda-result}
\Lambda(E) = \frac{\exp\!\big(\chi\frac{\arctan\sqrt{\energy}}{\sqrt{\energy}}\big)}{\abs{\Delta^+_0(E)}(1+\energy)}L(\energy)  \:,
\end{equation}
where the function $L$ can be related to $C$ with
\begin{equation}\label{eq:l-from-c}
L(\energy) = \frac{A}{\sqrt{2b}}\!\left[\frac{\anbohr}{2}\varkappa_0 + \gamma - \frac{C(\energy)\,(1+\energy)}{\chi\exp\!\big(\chi\frac{\arctan\sqrt{\energy}}{\sqrt{\energy}}\big)}\right]  \:.
\end{equation}
In contrast to $C$, the function $L$ is pretty close to a smoothly varying straight line, as shown in Fig.~\ref{fig:plot-overlap-linearized}.
Indeed, the three terms in the square brackets of Eq.~\eqref{eq:l-from-c} display very linear behaviors over a large energy range ($\lesssim 5~\mathrm{MeV}$).
This important feature is independent from our first guess~\eqref{eq:exp-fit-diproton} about the proton-proton wave function.
Therefore, the function $L$ is appropriate to model fitting over a large energy range.
\par Furthermore, the result~\eqref{eq:l-from-c} allows us to calculate the zero-energy values of $L$ and $\Lambda$.
Using $A=0.8850(5)~\mathrm{fm}^{-1/2}$, $\alpha_0=-7.815(9)~\mathrm{fm}$, and $C(0)=0.260(1)$ computed in potential models, we find
\begin{equation}\label{eq:l-zero-energy}
L(0) = \frac{A}{\sqrt{2b}}\left[\frac{-\anbohr}{2\alpha_0} + \gamma - \frac{C(0)}{\chi\E^{\chi}}\right] = 8.42(1)  \:.
\end{equation}
It should be noted that Bahcall and May originally obtained remarkably good estimates of $A$ and $C(0)$ from effective-range theory~\cite{Bahcall1969}
\begin{equation}\label{eq:bahcall-estimates}\begin{cases}
A    \simeq \sqrt{\frac{2b}{1-br_{\rm pn}}}    & = 0.883(2)~\mathrm{fm}^{-1/2}  \:,\\[10pt]
C(0) \simeq \frac{b(r_{\rm pp}+r_{\rm pn})}{4} & = 0.262(1)                     \:.
\end{cases}\end{equation}
In Eq.~\eqref{eq:bahcall-estimates}, we have used the numerical values of the proton-proton $^1S_0$ effective range $r_{\rm pp}=2.77(1)~\mathrm{fm}$ and the proton-neutron $^3S_1$ effective range $r_{\rm pn}=1.75(1)~\mathrm{fm}$ from Table XIV in Ref.~\cite{Machleidt2001}.
\par If, on the other hand, we perform the linear fitting of~$L$ on $[300,600]~\mathrm{keV}$ directly from Eqs.~\eqref{eq:weak-capture-lambda} and~\eqref{eq:lambda-result}, then we get
\begin{equation}\label{eq:linear-overlap}
L(\energy) = L_0 + L_1\energy = 8.42(1) + 0.55(1)\,\energy  \:.
\end{equation}%
\begin{figure}[ht]%
\inputpgf{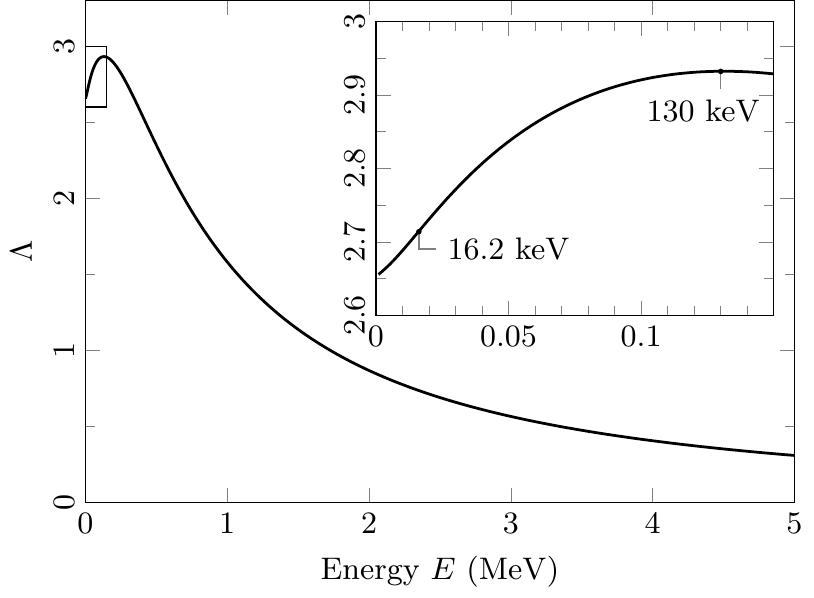}
\caption{Matrix element $\Lambda$ of the weak capture computed with the Reid93 potential.
The curve has an inflection point at $16.2~\mathrm{keV}$ and a maximum at $130~\mathrm{keV}$.
The inset is an enlargement of the graph to low energy.}
\label{fig:plot-overlap-lambda}
\end{figure}%
The uncertainties are due to the differences between the potential models.
The function $L$ of Eq.~\eqref{eq:linear-overlap} is compared to the results from potential models in Fig.~\ref{fig:plot-overlap-linearized}.
The actual curve of $L$ deviates from a straight line because of the short-range behavior of the wave functions.
Despite this deviation, it turns out that $L$ is accurate by less than $2\%$ error below $5~\mathrm{MeV}$, and is especially good below $1~\mathrm{MeV}$.
This adequacy confirms the validity of the parametrization~\eqref{eq:lambda-result}.
\par The matrix element $\Lambda$ computed in the Reid93 potential is depicted in Fig.~\ref{fig:plot-overlap-lambda}.
The curves obtained in other potentials, along with the result~\eqref{eq:lambda-result}, are indistinguishable at this scale.
These curves are quite rarely shown in the literature~\cite{Rupak2015}.
At zero energy, we find the important value for stellar nucleosynthesis $\Lambda^2(0)=7.034(33)$, which is consistent with Ref.~\cite{Adelberger2011}. 
Using Eq.~\eqref{eq:astrophysical-factor} and the numerical value of $\fermiint(Q)$ from Eq.~\eqref{eq:fermi-results} plus $1.62\%$ to account for radiative corrections~\cite{Adelberger2011, ChenJW2013, Marcucci2013, *Marcucci2014a, Kurylov2002, *Kurylov2003}, the corresponding value of $S(0)$ is $3.95(3)\times 10^{-23}~\mathrm{MeV}\,\mathrm{fm}^2$.

\section{Complex analysis of the weak capture\label{sec:complex-analysis}}
\par
\begin{figure}[ht]%
\inputpgf{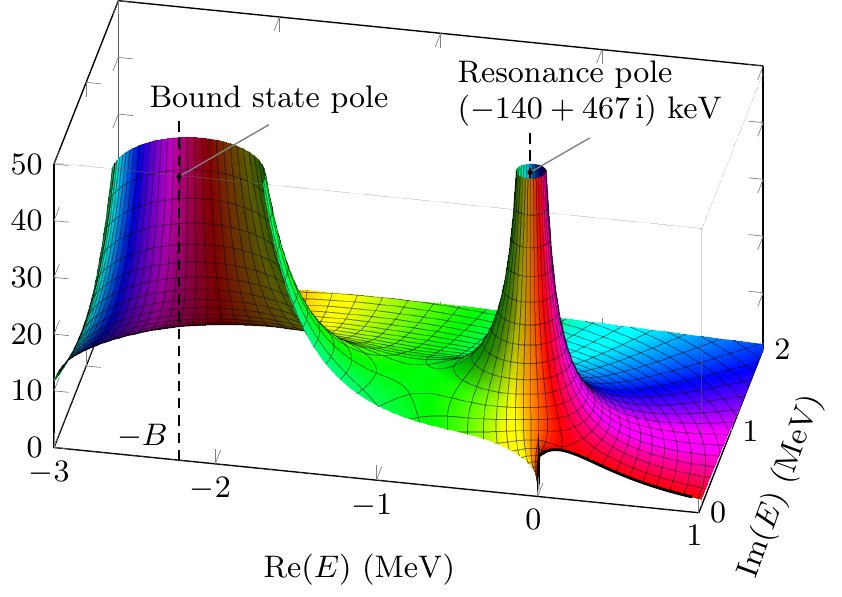}
\caption{3D plot of $\Lambda^2$ in the upper complex half-plane.
The vertical scale shows the modulus and the color highlights the phase.
The Coulomb singularities and the branch cut at $\arg(E)=\pi$ are not visible at this scale.}
\label{fig:plot3d-lambda2-high}%
\end{figure}

\begin{figure}[ht]%
\inputpgf{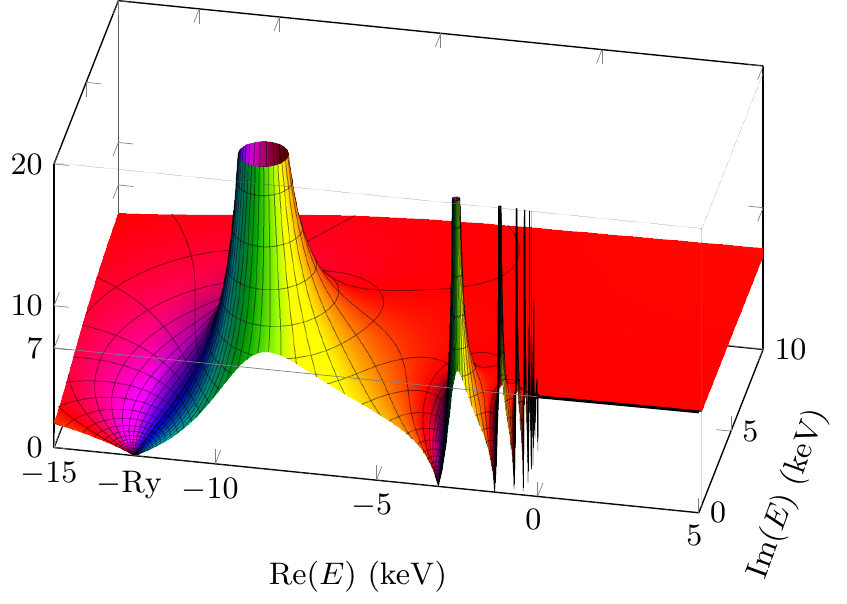}
\caption{Same as Fig.~\ref{fig:plot3d-lambda2-high} but for low energies.
The accumulation of alternating poles and zeros tends to $E=0$, and is thus an essential singularity.
The nuclear Rydberg energy is $\Ryd=12.4911~\mathrm{keV}$.}
\label{fig:plot3d-lambda2-low}
\end{figure}

 Remarkably, $\Lambda$ reaches a maximum near $130~\mathrm{keV}$, that corresponds, through Eq.~\eqref{eq:lambda-result}, to the minimum of $\abs{\Delta^+_0}$ seen in Fig.~\ref{fig:plot-erf}.
The actual origin of this maximum is revealed by the continuation of $\Lambda$ to complex energies, as provided by Eq.~\eqref{eq:lambda-result}.
The analytic continuation of $\Lambda^2$ based on Eqs.~\eqref{eq:lambda-result} and~\eqref{eq:linear-overlap} is shown in Figs.~\ref{fig:plot3d-lambda2-high} and~\ref{fig:plot3d-lambda2-low}.
Note that the curve along the positive real semi-axis in Fig.~\ref{fig:plot3d-lambda2-high} corresponds to Fig.~\ref{fig:plot-overlap-lambda}.
The deuteron bound state pole in Fig.~\ref{fig:plot3d-lambda2-high} is due to $(1+\energy)^{-1}$ in Eq.~\eqref{eq:lambda-result}.
\par
\begin{figure}[ht]%
\inputpgf{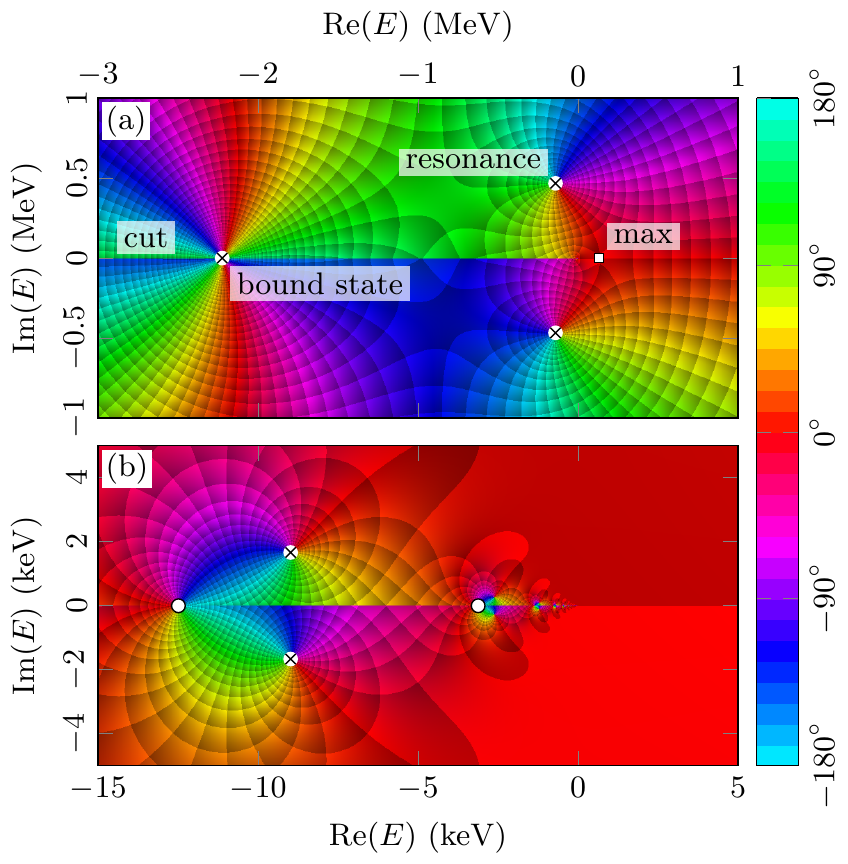}
\caption{Phase plots~\cite{GaspardD2018a, GaspardD2018b, Wegert2012} of the analytic continuation of $\Lambda^2$.
The complex phase is highlighted by colors.
The poles are marked by crosses, the zeros by empty circles, and the maximum at $0.13~\mathrm{MeV}$ by a square.
Panel (a) shows the high energies, and (b) is an enlargement into the low-energy region dominated by Coulomb singularities.
The branch cut along the negative real semi-axis stops at $E=0$.}
\label{fig:map-lambda2}
\end{figure}

 Furthermore, in contrast to $\Lambda$, the function $L$ is holomorphic in the neighborhood of $E=0$, because of the properties of the modified Coulomb functions $\Phi_{\eta,0}$ and $\Psi_{\eta,0}$ in the limit $E\rightarrow 0$~\cite{GaspardD2018a, GaspardD2018b}.
Therefore, it follows from Eq.~\eqref{eq:lambda-result} that any singularity of $\abs{\Delta^+_0}^{-1}$ is reflected on $\Lambda$.
In this regard, it can be shown that $\abs{\Delta^+_0}^{-1}$ has two poles at $\respole{\pm}$, that are interpreted as the proton-proton $^1S_0$ resonance poles~\cite{Kok1980, GaspardD2018a}.
In Fig.~\ref{fig:plot3d-lambda2-high}, only one of them is visible because the plot is restricted to the upper half-plane ($\Im E\geq0$).
The other one is shown in panel (a) of Fig.~\ref{fig:map-lambda2}.
\par The function $\abs{\Delta^+_0}^{-1}$ also possesses a branch cut along the negative real semi-axis due to the logarithm in the Bethe function~$g_0$ in Eq.~\eqref{eq:diproton-delta-plus-0}.
This branch cut makes $\Lambda$ complex at negative energy, although it is real at positive energy.
Being on the boundary of the plots, the branch cut cannot be seen either in Figs.~\ref{fig:plot3d-lambda2-high} or~\ref{fig:plot3d-lambda2-low}, but only in the top views of Fig.~\ref{fig:map-lambda2}.
\par In addition, $\abs{\Delta^+_0}^{-1}$ is responsible for the accumulation of poles and zeros shown in Fig.~\ref{fig:plot3d-lambda2-low}.
These singularities originate from the terms $\psi(\pm\I\eta)$ in the Bethe function~$g_0$.
The pole-zero pattern is repeated each $E=-\Ryd/n^2$ for $n\in\{1,2,3,\ldots\}$.
Such a structure can be understood as a set of virtual states generated by the Coulomb potential between the protons.
\par These considerations about the analytic properties of $\Lambda$ have two major consequences.
First, the maximum of $\Lambda$ is directly related to the Coulomb potential.
Especially, one sees in Fig.~\ref{fig:plot3d-lambda2-high} that it results from a saddle point between the conjugated resonance poles and the low-energy Coulomb singularities.
\par Second, $\Lambda$ is not analytic at $E=0$ due to $\Delta^+_0$ in Eq.~\eqref{eq:lambda-result}.
Therefore, its series expansion is not expected to converge over a nonzero energy range around $E=0$.
The common way of extracting the derivatives of $\Lambda$ would be to use polynomial extrapolation from data on a finite energy interval~\cite{Schiavilla1998, Marcucci2013, *Marcucci2014a, Acharya2016}.
However, such a method is not accurate due to the non-negligible influence of the interval itself~\cite{Acharya2016, Marcucci2013, *Marcucci2014a}.

\section{Zero-energy derivatives\label{sec:derivatives}}
One way to address the issue of the extrapolation to $E=0$ is to expand $\Lambda$ in power series directly from Eq.~\eqref{eq:lambda-result} taking advantage of the flatness of~$L$ at low energy.
The result~\eqref{eq:lambda-result} provides a constraint on the derivatives of $\Lambda$ with respect to the energy, especially the first derivative at zero energy: $\Lambda'(0)$.
This value plays a significant role in the proton-proton fusion at solar energies~\cite{Adelberger2011}.
From Eq.~\eqref{eq:lambda-result}, the logarithmic derivative of $\Lambda$ can be easily calculated
\begin{equation}\label{eq:lambda-log-der}
\der{\ln\Lambda}{E} = \der{\ln\phi_{1,0}}{E} - \der{\ln\Delta_0}{E} + \der{\ln L}{E}  \:,
\end{equation}
where the function $\abs{\Delta^+_0}$ has been replaced by $\Delta_0$ because they share the same asymptotic expansion at $E=0$.
This is due to the fact that the term $(\E^{2\pi\eta}-1)^{-2}$ in Eq.~\eqref{eq:diproton-delta-plus-2} is negligible since all of its derivatives are zero.
The first terms in the asymptotic expansion of $\Delta_0$ are
\begin{equation}
\Delta_0 = \frac{\anbohr}{2}\varkappa_0 - \left(\frac{E}{12\,\Ryd} + \frac{E^2}{120\,\Ryd^2} + \bigo(E^3)\right)  \:.
\end{equation}
However, it should be noted that this expansion does not converge at $E=0$ because of the Coulomb singularities in $\Delta_0$.
It remains nevertheless valid for $E\ll\Ryd$~\cite{GaspardD2018a, Baye2000a}.
In addition, the expansion of the regular Coulomb integral is given by
\begin{equation}
\phi_{1,0} = \chi\E^{\chi}\left(1 - \frac{3+\chi}{3}\energy + \frac{90+48\chi+5\chi^2}{90}\energy^2 - \bigo(\energy^3)\right)  \:.
\end{equation}
Combining these results in Eq.~\eqref{eq:lambda-log-der}, we find that the logarithmic derivative of $\Lambda$ at $E=0$ mostly depends on the effective-range parameters
\begin{equation}\label{eq:lambda-log-der-1}
\frac{\Lambda'(0)}{\Lambda(0)} = \frac{\abs{\alpha_0}m_{\rm p}}{2\hbar^2}\left(\frac{\anbohr}{3}-r_0\right) - \frac{3+\chi}{3B} + \frac{L_1}{BL_0}  \:.
\end{equation}
The prime over $\Lambda$ refers to the derivative with respect to~$E$.
This novel result was not obtained by Bahcall and May, although its numerical value is given in their paper~\cite{Bahcall1969}.
From $L_0$ and $L_1$ in Eq.~\eqref{eq:linear-overlap}, the result~\eqref{eq:lambda-log-der-1} yields
\begin{equation}\label{eq:lambda-der-1-exp-1}
\frac{\Lambda'(0)}{\Lambda(0)} = -0.4423(6) + (0.23149 - 0.01205r_0)\abs{\alpha_0}  \:.
\end{equation}
This result and all the following ones are expressed in units of MeV and fm.
\par It should be noted that, according to Eq.~\eqref{eq:l-from-c}, $L_0$ and $L_1$ also depend on effective-range parameters.
In this regard, the expansion~\eqref{eq:lambda-der-1-exp-1} is incomplete.
However, it is not possible to extract the full dependence of $L$ in the effective-range parameters since it would be necessary to modify the potential models accordingly.
\par Besides, one notices in Eq.~\eqref{eq:lambda-der-1-exp-1} that the uncertainty of the effective range $r_0$ only slightly influences the result.
Therefore, it is useful to re-express Eq.~\eqref{eq:lambda-der-1-exp-1} in the neighborhood of $\abs{\alpha_0}=7.815~\mathrm{fm}$ with $r_0=2.77(1)~\mathrm{fm}$.
We obtain the following result
\begin{equation}\label{eq:lambda-log-der-1-num}
\frac{\Lambda'(0)}{\Lambda(0)} = 1.106(2) + 0.1981(1)\:(\abs{\alpha_0}-7.815)  \:,
\end{equation}
around the scattering length $\alpha_0=-7.815(9)~\mathrm{fm}$.
The central value
\begin{equation}\label{eq:lambda-der-1-num}
\frac{\Lambda'(0)}{\Lambda(0)} = 1.106(3)~\mathrm{MeV}^{-1}  \:,
\end{equation}
is compatible with Refs.~\cite{Bahcall1969, ChenJW2013}.
It turns out that the term $L_1/L_0$ in Eq.~\eqref{eq:lambda-log-der-1}, which contains the short-range behavior of the wave functions, only contributes to about $2.6\%$.
Therefore, the uncertainty of $L_1/L_0$ marginally affects the overall uncertainty of $\Lambda'(0)/\Lambda(0)$, which is primarily due to the effective-range parameters $\alpha_0$ and $r_0$.
\par It is worth noting that Eq.~\eqref{eq:lambda-result} now determines all the derivatives of $\Lambda$ at $E=0$.
Indeed, the higher-order derivatives of $L$ are negligible in the expansion of $\Lambda$ compared to the other terms.
The reason is that the derivatives of $\Lambda$ are dominated by $\Delta^+_0$.
In particular, the second derivative of $\Lambda$ can also be calculated analytically from Eq.~\eqref{eq:lambda-result}.
We have
\begin{equation}\label{eq:lambda-der-2}
\begin{split}
\frac{\Lambda''(0)}{\Lambda(0)} & = \frac{\phi_{1,0}''}{B^2\phi_{1,0}} - \frac{\Delta_0''}{\Delta_0} + \frac{L_2}{B^2L_0} + 2\!\left(\frac{\Delta_0'}{\Delta_0}\right)^2  \\
 & - 2\frac{\phi_{1,0}'\Delta_0'}{B\phi_{1,0}\Delta_0} + 2\frac{\phi_{1,0}'L_1}{B^2\phi_{1,0}L_0} - 2\frac{L_1\Delta_0'}{BL_0\Delta_0}  \:,
\end{split}\end{equation}
where the primes refer to derivatives with respect to the main variable: either $E$ for $\Delta_0(E)$, or $\energy$ for $\phi_{1,0}(\energy)$.
All the functions in Eq.~\eqref{eq:lambda-der-2} are implicitly evaluated at $E=0$.
It turns out that the second derivative $L_2$ can be neglected, as it contributes to only $0.01\%$, far below the uncertainty of the other terms.
The terms of Eq.~\eqref{eq:lambda-der-2} containing $\Delta_0$ and its derivatives are dominating the other ones, especially the term $-\Delta_0''/\Delta_0$ which is about $29~\mathrm{MeV}^{-2}$.
Inserting the expansion of $\phi_{1,0}$ and $\Delta_0$ at $E=0$ into Eq.~\eqref{eq:lambda-der-2}, but without replacing the effective-range parameters $\alpha_0$ and $r_0$ by their numerical value for now, we find
\begin{equation}\label{eq:lambda-der-2-exp-1}
\begin{split}
\frac{\Lambda''(0)}{\Lambda(0)} & = 0.4051(5) + [3.5017(3) + 0.01066(1)r_0]\abs{\alpha_0}  \\
 & + \left(0.32737 - 0.01704r_0\right)^2 \abs{\alpha_0}^2   \:.
\end{split}\end{equation}
The last term in Eq.~\eqref{eq:lambda-der-2-exp-1} comes from the term $(\Delta_0'/\Delta_0)^2$ in Eq.~\eqref{eq:lambda-der-2}.
The uncertainties in this term are negligible, as it solely depends on accurately known physical quantities ($\hbar$, $\alpha$, and $m_{\rm p}$).
\par As previously, if we focus on the neighborhood of $\abs{\alpha_0}=7.815~\mathrm{fm}$ assuming $r_0=2.77(1)~\mathrm{fm}$, we get from Eq.~\eqref{eq:lambda-der-2-exp-1} the expression
\begin{equation}\label{eq:lambda-log-der-2}
\frac{\Lambda''(0)}{\Lambda(0)} = 32.795(6) + 4.758(2)\:(\abs{\alpha_0}-7.815)  \:.
\end{equation}
Note that the remainder term in Eq.~\eqref{eq:lambda-log-der-2} is $0.0785(1)\,(\abs{\alpha_0}-7.815)^2$ in unit $\mathrm{MeV}^{-2}$.
The central value
\begin{equation}\label{eq:lambda-der-2-num}
\frac{\Lambda''(0)}{\Lambda(0)} = 32.80(5)~\mathrm{MeV}^{-2}  \:,
\end{equation}
is in accordance with Ref.~\cite{ChenJW2013}.
The uncertainty on $\Lambda''(0)/\Lambda(0)$ is mainly due to $\alpha_0=-7.815(9)~\mathrm{fm}$.
This result~\eqref{eq:lambda-der-2-num} is considerably more accurate than what we get from direct fitting on $\Lambda$.
In fact, the direct computation of the second derivative of $\Lambda$ depends too much on the energy interval chosen for the fitting, hence degrading its accuracy.
Similar effective-range constraints can be derived for higher-order derivatives of~$\Lambda$ from Eq.~\eqref{eq:lambda-result}.
The expected accuracy of this approach does not exceed about $0.2\%$ as it is limited by the uncertainty on $\alpha_0$.
\par Finally, we deduce the zero-energy derivatives of the astrophysical $S$ factor from Eqs.~\eqref{eq:lambda-log-der-1-num} and~\eqref{eq:lambda-log-der-2}.
Knowing from Eq.~\eqref{eq:astrophysical-factor} that $S(E)$ is proportional to $\fermiint(E+Q)\abs{\Lambda(E)}^2$, the logarithmic derivatives read~\cite{Baye2013}
\begin{equation}
\frac{S'(0)}{S(0)} = \frac{\fermiint'(Q)}{\fermiint(Q)} + 2\frac{\Lambda'(0)}{\Lambda(0)}  \:,
\end{equation}
and
\begin{equation}
\frac{S''(0)}{S(0)} = \frac{\fermiint''(Q)}{\fermiint(Q)} + 4\frac{\fermiint'(Q)}{\fermiint(Q)}\frac{\Lambda'(0)}{\Lambda(0)} + 2\!\left(\frac{\Lambda'(0)}{\Lambda(0)}\right)^2 + 2\frac{\Lambda''(0)}{\Lambda(0)}  \:.
\end{equation}
Using the values of the Fermi phase-space integral from Eq.~\eqref{eq:fermi-results}, we get the results
\begin{equation}\label{eq:astro-log-der}
\begin{aligned}
& \frac{S'(0)}{S(0)}  = 11.253(3) + 0.3962(2) \:(\abs{\alpha_0}-7.815)  ,\\
& \frac{S''(0)}{S(0)} = 169.51(8) + 17.56(1)  \:(\abs{\alpha_0}-7.815)  .
\end{aligned}\end{equation}
The central values $11.25(1)~\mathrm{MeV}^{-1}$ and $169.5(3)~\mathrm{MeV}^{-2}$, obtained by setting $\abs{\alpha_0}=7.815(9)~\mathrm{fm}$, are compatible with Refs.~\cite{Bahcall1969, Adelberger2011, ChenJW2013, Marcucci2013, *Marcucci2014a, Angulo1999, Baye2013}.
These results are obtained with an unprecedented high accuracy.
In the literature, most of the uncertainties are due to the polynomial extrapolation of $S(E)$ which is highly sensitive to the chosen energy interval~\cite{Marcucci2013, *Marcucci2014a, Acharya2016}.
Our method is based instead on the fitting of~$L$, as suggested by the analytic structure of~$\Lambda$ at low energy.
Consequently, the results~\eqref{eq:astro-log-der} are not affected by the uncertainty of $1\%$ reported for $S(0)$~\cite{Adelberger2011}.

\section{Conclusion\label{sec:conclusion}}
To conclude, we have derived an accurate parametrization of the energy dependence of the weak capture matrix element $\Lambda$ valid up to a few MeVs, that is based on recent effective-range functions~\cite{RamirezSuarez2017, GaspardD2018a}.
This result provides the analytic continuation of $\Lambda$ to complex energies, and highlights the relationship between its maximum near $0.13~\mathrm{MeV}$, the broad proton-proton resonance, and the Coulomb sub-threshold singularities.
In addition, it leads to a remarkably accurate determination of the logarithmic derivatives of the astrophysical $S$ factor at $E=0$ in terms of effective-range parameters.
Our method bypasses the issue~\cite{Acharya2016} of the energy-range dependence in the polynomial fitting of $S$ by means of the function $L$, that is analytic at low energy, in contrast to~$S$.
In this regard, the gain in accuracy on $S(0)$, $S'(0)$, and $S''(0)$ using our method is expected to be similar if corrections, such as the two-body current terms~\cite{Schiavilla1998, Park2003, Marcucci2013, *Marcucci2014a}, are taken into account.
Finally, the new parametrization~\eqref{eq:lambda-result} is appropriate for use in stellar and Big-Bang astrophysics as it covers a large energy range up to the binding energy of the deuteron.
\begin{acknowledgments}
This work was supported by the European Union's \href{https://doi.org/10.13039/100010662}{Horizon 2020 (Excellent Science)} research and innovation program under Grant Agreement No. 654002.
\end{acknowledgments}%
\appendix
\section{Fermi phase-space integral\label{app:fermi-integral}}
When calculating the proton-proton weak capture cross section, we are led to integrate the Dirac delta of energy-momentum conservation over the momenta of the three outgoing particles: the deuteron, the positron, and the electronic neutrino.
The resulting integral is known as the Fermi phase-space integral and reads in first approximation~\cite{Bahcall1966a, Wilkinson1982, ChenJW2013}
\begin{equation}\label{eq:fermi-int}
\fermiint(E+Q) = \int_1^{\wmax} P(w)\,w\sqrt{w^2-1}\,(\wmax-w)^2\D w  \:,
\end{equation}
as long as the recoil of the deuteron is neglected.
The released energy $Q=2m_{\rm p}c^2-m_{\rm d}c^2-m_{\rm e}c^2$ is found to be $0.420236(17)~\mathrm{MeV}$ with the masses from Ref.~\cite{Tanabashi2018}.
The variable $w$ in Eq.~\eqref{eq:fermi-int} is the positron energy divided by its mass.
With this notation, $\sqrt{w^2-1}$ is to be understood as the positron momentum divided by its mass.
From energy conservation, the upper bound denoted as $\wmax$ is equal to $(E+Q+m_{\rm e}c^2)/(m_{\rm e}c^2)$.
It means that the Fermi integral $\fermiint$ also depends on the proton-proton energy $E$.
The purpose of this Appendix is to calculate the low-energy dependence of $\fermiint$ on $E$.
\par The Coulomb factor $P$ in Eq.~\eqref{eq:fermi-int}, accounting for the distortion of the positron wave function in the electric field of the deuteron, is given by~\cite{Bahcall1966a, Wilkinson1982, ChenJW2013}
\begin{equation}\label{eq:fermi-coulomb}
P(w) = 2(1+\nu)\left(2\rho\sqrt{w^2-1}\right)^{-2(1-\nu)}\frac{\abs{\Gamma(\nu+\I\eta_{\rm e})}^2}{\E^{\pi\eta_{\rm e}}\Gamma(2\nu+1)^2}  \:,
\end{equation}
where $\nu$ is equal to $\sqrt{1-\alpha^2}$ with the fine-structure constant $\alpha\simeq 1/137.036$, and $\rho=Rm_{\rm e}c^2/(\hbar c)$ is the dimensionless radius of the deuteron.
In the following calculations, we will assume $R=2.14~\mathrm{fm}$~\cite{ChenJW2013}.
In Eq.~\eqref{eq:fermi-coulomb}, the Sommerfeld parameter of the emitted positron $\eta_{\rm e}=\alpha w/\sqrt{w^2-1}$ must be positive, as it is repelled by the nucleus.
Conversely, in a $\beta^-$ decay, the Sommerfeld parameter $\eta_{\rm e}$ should take a minus sign.
It should be noted that in the nonrelativistic limit ($\nu\rightarrow 1$), the Coulomb distortion factor $P(w)$ becomes
\begin{equation}
P(w) = \abs{C_{\eta_{\rm e},0}}^2 = \frac{2\pi\eta_{\rm e}}{\E^{2\pi\eta_{\rm e}} - 1}  \:.
\end{equation}
\par The Fermi integral~\eqref{eq:fermi-int} cannot be analytically calculated in a simple form.
However, very efficient approximations exist.
One way is to expand the Coulomb factor~\eqref{eq:fermi-coulomb} in series of the fine-structure constant $\alpha$.
We find
\begin{equation}\label{eq:fermi-coulomb-series}\begin{split}
P(w) & = 1 - \frac{\alpha\pi w}{\sqrt{w^2-1}} + \alpha^2\bigg[\frac{\pi^2}{3}\bigg(\frac{w}{\sqrt{w^2-1}}\bigg)^2 \bigg. \\
 & + \bigg. \frac{11}{4} - \gamma - \ln\!\big(2\rho\sqrt{w^2-1}\big)\bigg] + \bigo(\alpha^3)  \:,
\end{split}\end{equation}
where $\gamma=0.5772\ldots$ is the Euler-Mascheroni constant.
In this work, we limit ourselves to the order $\alpha^2$, as it is enough to obtain at least five decimal places in the final results.
The same approach is followed in Ref.~\cite{Wilkinson1982} up to $\alpha^3$.
Now, we just have to calculate one Fermi integral for each term in the expansion~\eqref{eq:fermi-coulomb-series}.
The advantage is that the integrals of the form
\begin{equation}\label{eq:fermi-terms}
f_p(\wmax) = \int_1^{\wmax} \left(\frac{w}{\sqrt{w^2-1}}\right)^p w\sqrt{w^2-1} (\wmax-w)^2\D w  \:,
\end{equation}
which will come into play, can be expressed in terms of elementary functions for $p\in\mathbb{Z}$.
Such expressions can be obtained by expanding the last factor $(\wmax-w)^2$ in Eq.~\eqref{eq:fermi-terms}.
The results read for $p=0$
\begin{equation}\label{eq:fermi-term-0}\begin{split}
f_0(\wmax) & = \left(\frac{\wmax^4}{30} - \frac{3\wmax^2}{20} - \frac{2}{15}\right)\sqrt{\wmax^2-1}  \\
 & + \frac{\wmax}{4}\ln\!\left(\wmax + \sqrt{\wmax^2-1}\right)  \:,
\end{split}\end{equation}
for $p=1$
\begin{equation}\label{eq:fermi-term-1}
f_1(\wmax) = \frac{\wmax^5}{30} - \frac{\wmax^2}{3} + \frac{\wmax}{2} - \frac{1}{5}  \:,
\end{equation}
and for $p=2$
\begin{equation}\label{eq:fermi-term-2}\begin{split}
f_2(\wmax) & = \left(\frac{\wmax^4}{30} + \frac{11\wmax^2}{60} + \frac{8}{15}\right)\sqrt{\wmax^2-1}  \\
 & - \frac{3\wmax}{4}\ln\!\left(\wmax + \sqrt{\wmax^2-1}\right)   \:.
\end{split}\end{equation}
We notice that, according to Eqs.~\eqref{eq:fermi-term-0}, \eqref{eq:fermi-term-1}, and~\eqref{eq:fermi-term-2}, the Fermi integral is expected to behave as $\bigo(E^5)$ at relatively large energy ($E\gg m_{\rm e}c^2$).
Therefore, $\fermiint(E+Q)$ dominates the $\bigo(E^{-2})$ behavior of $\Lambda^2$ in Eq.~\eqref{eq:astrophysical-factor}.
\par The factor $\sqrt{w^2-1}$ in the logarithmic term of expansion~\eqref{eq:fermi-coulomb-series} can be neglected because it remains of the order of $1$ except at large proton-proton energies ($E\gg m_{\rm e}c^2$).
Therefore, using Eq.~\eqref{eq:fermi-terms}, the Fermi integral~\eqref{eq:fermi-int} is approximated by
\begin{equation}\label{eq:fermi-approx}\begin{split}
 & \fermiint(E+Q) \simeq f_0(\wmax) - \alpha\pi f_1(\wmax)   \\
 & + \alpha^2\left[\frac{\pi^2}{3}f_2(\wmax) + \left(\frac{11}{4} - \gamma - \ln(2\rho)\right)f_0(\wmax)\right]  \:.
\end{split}\end{equation}
This expression allows us to find at least five decimal places without requiring numerical integration.
Another advantage is the computation of the derivatives of the Fermi integrals with respect to $E$.
In this work, we need the first two derivatives of $\fermiint(E+Q)$ at zero proton energy ($E=0$).
This can be easily achieved with the derivatives of $f_p(\wmax)$ with respect to $\wmax$ that are obtained directly from Eqs.~\eqref{eq:fermi-term-0}, \eqref{eq:fermi-term-1}, and~\eqref{eq:fermi-term-2}.
The derivatives of $\fermiint(E+Q)$ with respect to $E$ have thus essentially the same expressions as Eq.~\eqref{eq:fermi-approx} by replacing $f_p(\wmax)$ with the derivatives with respect to $\wmax$, denoted as $f_p^{(n)}(\wmax)$.
Note the change of variable $\partial_E^n\fermiint = (m_{\rm e}c^2)^{-n}\partial_{\wmax}^n\fermiint$ in the manipulation.
\begin{table}[ht]%
\centering\begin{tabular}{l*{3}{c}}\hline
            & $n=0$        & $n=1$        & $n=2$       \\ \hline
$f_0^{(n)}$ & $0.14827(2)$ & $0.68187(8)$ & $2.3574(2)$ \\
$f_1^{(n)}$ & $0.27417(4)$ & $1.1233(1)$  & $3.3682(2)$ \\
$f_2^{(n)}$ & $0.64955(7)$ & $2.2505(2)$  & $5.4045(3)$ \\ \hline
\end{tabular}%
\caption{Numerical values of the functions $f_p(\wmax)$ and their derivatives with respect to $\wmax$ at $E=0$.
The upper index $n$ is the order of the derivatives.}
\label{tab:fermi-terms}
\end{table}%
The numerical values of the functions $f_p(\wmax)$ and their derivatives at $E=0$, that is for $\wmax=(Q+m_{\rm e}c^2)/(m_{\rm e}c^2)=1.82238(3)$, are given in Table~\ref{tab:fermi-terms}.
Inserting the numerical values of Table~\ref{tab:fermi-terms} in the approximation~\eqref{eq:fermi-approx} for the different derivative orders ($n=0,1,2$) leads to the results
\begin{equation}\label{eq:fermi-results}\begin{cases}
\fermiint(Q)                 = 0.14215(2)                   \:,\\
\fermiint'(Q)/\fermiint(Q)   = 9.0413(3)~\mathrm{MeV}^{-1}  \:,\\
\fermiint''(Q)/\fermiint(Q)  = 61.479(5)~\mathrm{MeV}^{-2}  \:.
\end{cases}\end{equation}
These results have also been checked by numerical integration in Wolfram Mathematica~\cite{Wolfram1999}.
The uncertainties in Table~\ref{tab:fermi-terms} and Eq.~\eqref{eq:fermi-results} come from the released energy $Q$. 
Finally, the low-energy behavior of the Fermi integral can be written as
\begin{equation}
\frac{\fermiint(E+Q)}{\fermiint(Q)} = 1 + \frac{\fermiint'(Q)}{\fermiint(Q)}E + \frac{\fermiint''(Q)}{\fermiint(Q)}\frac{E^2}{2} + \bigo(E^3)  \:,
\end{equation}
with the numerical values of Eq.~\eqref{eq:fermi-results}.
\par It should be noted that the third derivative of the Fermi function~\eqref{eq:fermi-int} with respect to $E$ is devoid of integral and can be expressed exactly in terms of $P(\wmax)$.
We have
\begin{equation}
\der[3]{\fermiint}{E}(E+Q) = \frac{2}{(m_{\rm e}c^2)^3}P(\wmax)\,\wmax\sqrt{\wmax^2-1}  \:,
\end{equation}
from which the numerical value $\fermiint^{(3)}(Q) = 40.498(2)~\mathrm{MeV}^{-3}$ at $E=0$ is easily found.
Our approach avoids using numerical derivatives, as they are ill-conditioned in finite precision arithmetic, especially for high-order derivatives.
This also ensures the accuracy of the results~\eqref{eq:fermi-results}.

\section{Laplace transforms of the modified Coulomb functions\label{app:coulomb-integrals}}
In this Appendix, we present the derivation of the Laplace transforms of the modified Coulomb wave functions $\Phi_{\eta,\ell}(\rho)$ and $\Psi_{\eta,\ell}(\rho)$ defined in Ref.~\cite{GaspardD2018b}.
More explicitly, we are looking for analytical expressions of the integrals
\begin{equation}\label{eq:integral-phi}
\phi_{\beta,\ell}(\energy) = \int_0^\infty \E^{-\beta x}\,\Phi_{\eta,\ell}(x\sqrt{\energy}) \D x  \:,
\end{equation}
and
\begin{equation}\label{eq:integral-psi}
\psi_{\beta,\ell}(\energy) = \int_0^\infty \E^{-\beta x}\,\Psi_{\eta,\ell}(x\sqrt{\energy}) \D x  \:,
\end{equation}
where $x=br$ is the dimensionless radial coordinate, and $\sqrt{\energy}=k/b$ is the dimensionless wave number.
Although we only need the result for $\ell=0$, we have made our derivation more general.
The reason is that we use the connection formula between $\Phi_{\eta,\ell}(\rho)$ and $\Psi_{\eta,\ell}(\rho)$ developed in Ref.~\cite{GaspardD2018b} to calculate $\psi_{\beta,\ell}(\energy)$ on the basis of $\phi_{\beta,\ell}(\energy)$ for any integer $\ell$.

\subsection{Regular Coulomb integral}
As shown in Ref.~\cite{GaspardD2018b}, the Coulomb function $\Phi_{\eta,\ell}$ in Eq.~\eqref{eq:integral-phi} is given by
\begin{equation}\label{eq:coulomb-phi-def}
\Phi_{\eta,\ell}(\kappa x) = (\chi x)^{\ell+1}\E^{\I\kappa x} \regm{\ell+1+\I\eta}{2\ell+2}{-2\I\kappa x}  \:,
\end{equation}
where $\kappa=\sqrt{\energy}$, and $\chi=2/(\anbohr b)$ is the Bahcall and May constant~\cite{Bahcall1969}.
The regularized confluent hypergeometric function in Eq.~\eqref{eq:coulomb-phi-def} is defined by the series~\cite{Olver2010}
\begin{equation}\label{eq:1f1-def}
\regm{a}{b}{z} = \frac{1}{\Gamma(b)}\hypm{a}{b}{z} = \sum_{n=0}^\infty \frac{(a)_n}{\Gamma(b+n)} \frac{z^n}{n!}  \:,
\end{equation}
where $(a)_n=\Gamma(a+n)/\Gamma(a)$ is the Pochhammer symbol.
The division by $\Gamma(b)$ in Eq.~\eqref{eq:1f1-def} eliminates the singularities of $\hypm{a}{b}{z}$ at $b\in\mathbb{Z}_{\leq0}$~\cite{Olver2010}.
Using the definition~\eqref{eq:coulomb-phi-def}, the regular Coulomb integral~\eqref{eq:integral-phi} expands as follows
\begin{equation}\label{eq:integral-phi-step-1}\begin{split}
\phi_{\beta,\ell} & = \chi^{\ell+1}\sum_{n=0}^\infty \frac{(\ell+1+\I\eta)_n(-2\I\kappa)^n}{\Gamma(2\ell+2+n)\,n!}  \\
 & \times\int_0^\infty \E^{-(\beta-\I\kappa)x}x^{n+\ell+1}\D x  \:.
\end{split}\end{equation}
All the remaining integrals in Eq.~\eqref{eq:integral-phi-step-1} are given by
\begin{equation}\label{eq:integral-phi-step-2}
\int_0^\infty \E^{-(\beta-\I\kappa)x}x^{n+\ell+1}\D x = \frac{\Gamma(n+\ell+2)}{(\beta-\I\kappa)^{n+\ell+2}}  \:.
\end{equation}
One notices that the combination of Eqs.~\eqref{eq:integral-phi-step-1} and~\eqref{eq:integral-phi-step-2} leads to the Gauss hypergeometric function ${}_2F_1$, or more specifically to its regularized version~\cite{Olver2010}
\begin{equation}\label{eq:2f1-def}
\regf{a}{b}{c}{z} = \frac{1}{\Gamma(c)}\hypf{a}{b}{c}{z} = \sum_{n=0}^\infty \frac{(a)_n(b)_n}{\Gamma(c+n)} \frac{z^n}{n!}  \:.
\end{equation}
Using the definition~\eqref{eq:2f1-def} in Eq.~\eqref{eq:integral-phi-step-1}, we obtain the following result:
\begin{equation}\label{eq:integral-phi-full}
\phi_{\beta,\ell}(\energy) = \frac{\chi^{\ell+1}\Gamma(\ell+2)}{(\beta-\I\kappa)^{\ell+2}} \regf{\ell+2}{\ell+1+\I\eta}{2\ell+2}{\tfrac{-2\I\kappa}{\beta-\I\kappa}}  \:.
\end{equation}
Remarkably, this result is considerably simplified in the special case $\ell=0$.
Indeed, the hypergeometric function in Eq.~\eqref{eq:integral-phi-full} is then of the form $\hypf{a}{b}{a}{z}$, which reduces to $(1-z)^{-b}$~\cite{Olver2010} because of the simplification in the series~\eqref{eq:2f1-def}.
From Eq.~\eqref{eq:integral-phi-full}, one finds
\begin{equation}\label{eq:integral-phi-0}
\phi_{\beta,0}(\energy) = \frac{\chi}{\beta^2+\kappa^2} \E^{2\eta\arctan(\kappa/\beta)}   \:.
\end{equation}
This useful result is at the basis of the parametrization of $\Lambda(E)$ proposed in this paper.

\subsection{Irregular Coulomb integral}
Now, we calculate the Laplace transform~\eqref{eq:integral-psi} of $\Psi_{\eta,\ell}$.
This calculation is significantly less straightforward than for $\Phi_{\eta,\ell}$, because it does not reduce to elementary functions for $\ell=0$.
The Coulomb function $\Psi_{\eta,\ell}$ is defined in Ref.~\cite{GaspardD2018b} as
\begin{equation}\label{eq:coulomb-psi-def}\begin{split}
\Psi_{\eta,\ell}(\rho) & = w_{\eta\ell}\Gamma(-\ell\pm\I\eta)(2\eta\rho)^{\ell+1}\E^{\pm\I\rho}\hypu{\ell+1\pm\I\eta}{2\ell+2}{\mp2\I\rho}   \\
 & - w_{\eta\ell}h^\pm_{\eta\ell}\Phi_{\eta,\ell}(\rho)  \:,
\end{split}\end{equation}
where the choice of the upper or lower sign is immaterial.
In Eq.~\eqref{eq:coulomb-psi-def}, $\hypu{a}{b}{z}$ is the Tricomi confluent hypergeometric function, and the Bethe functions $h^\pm_{\eta\ell}$ are defined as
\begin{equation}\label{eq:bethe-h}
h^\pm_{\eta\ell} = \frac{\psi(\ell+1\pm\I\eta)+\psi(-\ell\pm\I\eta)}{2} - \ln(\pm\I\eta)  \:.
\end{equation}
The subtraction by $w_{\eta\ell}h^\pm_{\eta\ell}\Phi_{\eta,\ell}$ in Eq.~\eqref{eq:coulomb-psi-def} is intended to compensate for the singularities of $\hypu{a}{b}{z}$ in the complex plane of the energy.
This operation makes $\Psi_{\eta,\ell}(kr)$ regular for $k\in\mathbb{C}$~\cite{GaspardD2018b}.
\par Performing the direct integration of Eq.~\eqref{eq:coulomb-psi-def} by means of the integral representation of $\hypu{a}{b}{z}$~\cite{Olver2010} leads to
\begin{equation}\label{eq:integral-psi-full}\begin{split}
\psi_{\beta,\ell} & = \tfrac{\Gamma(\ell+1\pm\I\eta)\Gamma(\ell+2)\Gamma(1-\ell)}{\mp2\I\kappa(\pm\I\eta)^\ell}\regf{\ell+2}{1-\ell}{2\pm\I\eta}{\tfrac{\beta\pm\I\kappa}{\pm 2\I\kappa}}  \\
 & - w_{\eta\ell}h^\pm_{\eta\ell}\phi_{\beta,\ell}  \:.
\end{split}\end{equation}
Note that, this function is not finite for partial waves higher than $S$ ($\ell>0$) due to the vertical asymptote of $\Psi_{\eta,\ell}(kr)$ at $r=0$.
When $\ell=0$, the hypergeometric function in the above equation can be efficiently computed from its continued fraction expansion.
\par The expression~\eqref{eq:integral-psi-full} is quite difficult to analyze at low energy because the hypergeometric function shows an essential singularity at $\energy=0$.
Although this singularity is compensated by $w_{\eta\ell}h^\pm_{\eta\ell}\phi_{\beta,\ell}$, it prevents the hypergeometric function from having a convergent low-energy expansion.
This is why we propose to determine a suitable approximation to Eq.~\eqref{eq:integral-psi-full} from another approach.
\par It turns out that the function $\Psi_{\eta,\ell}$ is related to the regular Coulomb function $\Phi_{\eta,\ell}$.
We have shown in Ref.~\cite{GaspardD2018b} that $\Psi_{\eta,\ell}$ obeys the following connection formula
\begin{equation}\label{eq:coulomb-psi-from-phi}
\Psi_{\eta,\ell} = \frac{w_{\eta\ell}}{2}\dot{\Phi}_{\eta,\ell} + \frac{1}{2}\dot{\Phi}_{\eta,-\ell-1}  \:,
\end{equation}
where the dots refer to derivatives with respect to $\ell$.
This useful property is preserved by the Laplace transforms~\eqref{eq:integral-phi} and~\eqref{eq:integral-psi}.
Therefore, the function $\psi_{\beta,\ell}(\energy)$ can be calculated from derivatives of $\phi_{\beta,\ell}(\energy)$ as follows~\cite{GaspardD2018b}
\begin{equation}\label{eq:integral-psi-from-phi}
\frac{\psi_{\beta,\ell}}{w_{\eta\ell}\phi_{\beta,\ell}} = \frac{1}{2}\left(\frac{\dot{\phi}_{\beta,\ell}}{\phi_{\beta,\ell}} + \frac{\dot{\phi}_{\beta,-\ell-1}}{\phi_{\beta,-\ell-1}}\right)  \:.
\end{equation}
However, in order to calculate the derivatives in Eq.~\eqref{eq:integral-psi-from-phi}, we need to use the general expression~\eqref{eq:integral-phi-full} valid of $\phi_{\beta,\ell}$ for all $\ell\in\mathbb{C}$.
In this regard, we have found convenient to approximate the hypergeometric function by its low-energy confluent limit
\begin{equation}\label{eq:2f1-low-energy-approx}
\regf{\ell+2}{\ell+1+\I\eta}{2\ell+2}{\tfrac{-2\I\kappa}{\beta-\I\kappa}} = \regm{\ell+2}{2\ell+2}{\tfrac{\chi}{\beta}} + \bigo(\kappa^2)  \:.
\end{equation}
The approximation~\eqref{eq:2f1-low-energy-approx} could be improved at $\kappa=0$ by the confluence expansion (20a) in Ref.~\cite{Nagel2004}.
However, this expansion converges so slowly for $\kappa>1$ that we will not use it here.
The advantage of the approximation~\eqref{eq:2f1-low-energy-approx} is the consistency with Bahcall's and May's results in the zero-energy limit.
It is useful in our calculation to rewrite the confluent hypergeometric function in Eq.~\eqref{eq:2f1-low-energy-approx} in terms of the regular Coulomb function
\begin{equation}\label{eq:1f1-from-phi}
\regm{\ell+2}{2\ell+2}{\tfrac{\chi}{\beta}} = (\beta/\chi)^{\ell+1}\E^{\chi/2\beta}\Phi_{-\I,\ell}(\I\chi/2\beta)  \:.
\end{equation}
The approximation of $\phi_{\beta,\ell}$ for $\ell\ll1$ is thus given by
\begin{equation}\label{eq:integral-phi-approx}
\phi_{\beta,\ell} \simeq \frac{\beta^{\ell+1}\Gamma(\ell+2)}{\sqrt{\beta^2+\kappa^2}^{\ell+2}} \E^{2\eta\arctan(\kappa/\beta)} \E^{-\chi/2\beta}\Phi_{-\I,\ell}(\I\chi/2\beta)  \:.
\end{equation}
From Eq.~\eqref{eq:integral-phi-full} to Eq.~\eqref{eq:integral-phi-approx}, we have taken the modulus of the factor $(\beta-\I\kappa)^{-\ell-2}$ because $\phi_{\beta,\ell}$ should still remain a positive real function after the approximation~\eqref{eq:2f1-low-energy-approx}.
The logarithmic derivative of $\phi_{\beta,\ell}$ with respect to $\ell$ can be easily calculated from Eq.~\eqref{eq:integral-phi-approx}:
\begin{equation}\label{eq:dot-integral-phi}
\frac{\dot{\phi}_{\beta,\ell}}{\phi_{\beta,\ell}} \simeq \psi(\ell+2) - \ln\sqrt{1+\kappa^2/\beta^2} + \frac{\dot{\Phi}_{-\I,\ell}(\I\chi/2\beta)}{\Phi_{-\I,\ell}(\I\chi/2\beta)}  \:.
\end{equation}
We neglect the logarithmic term in Eq.~\eqref{eq:dot-integral-phi} because it is irrelevant in the $\bigo(\kappa^2)$ approximation of Eq.~\eqref{eq:2f1-low-energy-approx}.
Combining two expressions~\eqref{eq:dot-integral-phi} evaluated at $\ell$ and $-\ell-1$ in Eq.~\eqref{eq:integral-psi-from-phi}, we get
\begin{equation}\label{eq:integral-psi-approx-step-1}
\frac{\psi_{\beta,\ell}}{w_{\eta\ell}\phi_{\beta,\ell}} \simeq \frac{\psi(\ell+2) + \psi(1-\ell)}{2} + \frac{\Psi_{-\I,\ell}(\I\chi/2\beta)}{w_{-\I,\ell}\Phi_{-\I,\ell}(\I\chi/2\beta)}  \:.
\end{equation}
This relation can be simplified further by means of the definition~\eqref{eq:coulomb-psi-def} of $\Psi_{\eta,\ell}$ with the plus sign.
Finally, after the elimination of the digamma functions with Eq.~\eqref{eq:bethe-h}, we obtain
\begin{equation}\label{eq:integral-psi-approx-step-2}
\frac{\psi_{\beta,\ell}}{w_{\eta\ell}\phi_{\beta,\ell}} \simeq \Gamma(1-\ell)\frac{\hypu{\ell+2}{2\ell+2}{\chi/\beta}}{\regm{\ell+2}{2\ell+2}{\chi/\beta}}  \:.
\end{equation}
In the special case of interest $\ell=0$, this result can be written as
\begin{equation}\label{eq:integral-psi-approx-0}
\frac{\psi_{\beta,0}}{\phi_{\beta,0}} \simeq \Gamma(-1,\chi/\beta)  \:,
\end{equation}
where $\Gamma(a,z)$ is the upper incomplete gamma function defined by
\begin{equation}\label{eq:upper-incomplete-gamma}
\Gamma(a,z) = \int_z^\infty t^{a-1}\E^{-t} \D t  \:.
\end{equation}%
\begin{figure}[ht]%
\inputpgf{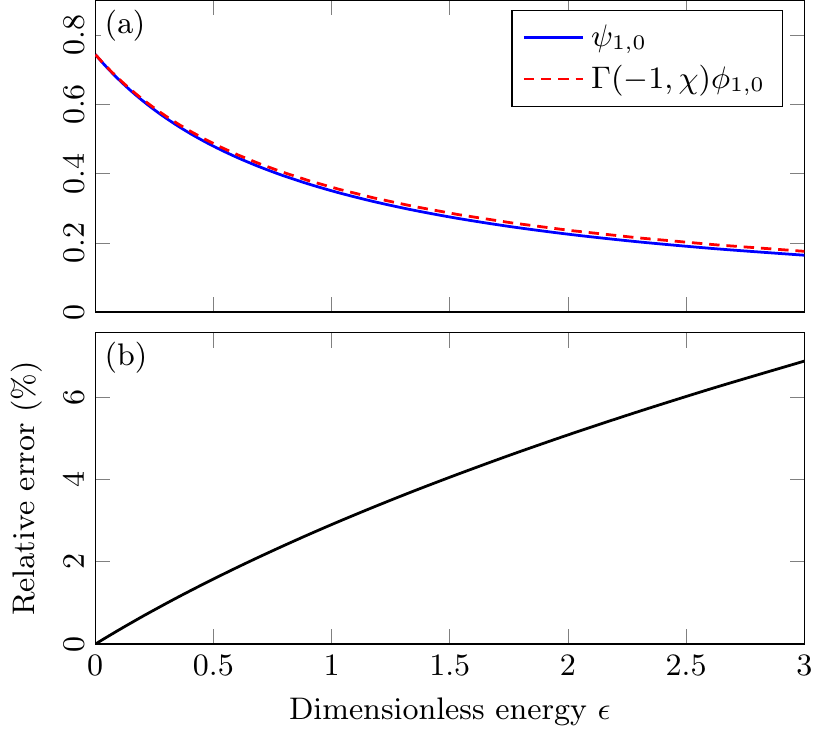}%
\caption{Comparison between the Coulomb integrals $\phi_{\beta,\ell}(\energy)$ and $\psi_{\beta,\ell}(\energy)$ for $\beta=1$, $\chi=2/(\anbohr b)=0.149816$, and $\ell=0$.
Panel (a) shows $\psi_{1,0}(\energy)$ and $\Gamma(-1,\chi)\phi_{1,0}(\energy)$, and panel (b) depicts the relative error between $\psi_{1,0}$ computed from Eq.~\eqref{eq:integral-psi-full} and the approximation~\eqref{eq:integral-psi-approx-0}.}
\label{fig:plot-coulomb-integrals}
\end{figure}%
When $\beta=1$ and $\chi=0.149816$, the ratio~\eqref{eq:integral-psi-approx-0} evaluates to about $4.28065$.
The incomplete gamma function in Eq.~\eqref{eq:integral-psi-approx-0} can also be related to the exponential integral $E_1(z)=\Gamma(0,z)$ as done in Bahcall's and May's work~\cite{Bahcall1969}:
\begin{equation}
\Gamma(-1,z) = \frac{\E^{-z}}{z} - E_1(z)  \:.
\end{equation}%
Bahcall, however, limited his calculation to zero energy, in contrast to the property~\eqref{eq:integral-psi-approx-0} valid up to a few MeVs.
\par Furthermore, the novel result~\eqref{eq:integral-psi-approx-0} means that $\psi_{\beta,0}$ is nearly proportional to $\phi_{\beta,0}$ on a large energy range.
The accuracy of this property is graphically tested in Fig.~\ref{fig:plot-coulomb-integrals}.
As we can see, the relative error of the estimate at $\energy=1$, that corresponds to $E=B$, is only $2.9\%$.
The overall accuracy of the approximation~\eqref{eq:integral-psi-approx-0} over a few MeVs is primarily due to the smallness of $\chi$ with respect to $1$ ($\chi=0.149816$).
In fact, it can be shown that both $\psi_{\beta,0}$ and $\Gamma(-1,\chi/\beta)\phi_{\beta,0}$ have the same neutral-charge limit:
\begin{equation}
\lim_{\chi\rightarrow 0} \psi_{\beta,0} = \lim_{\chi\rightarrow 0} \Gamma(-1,\chi/\beta)\phi_{\beta,0} = \frac{\beta}{\beta^2+\energy}  \:.
\end{equation}
Therefore, the property~\eqref{eq:integral-psi-approx-0} tends to be exact for $\chi\rightarrow 0$, but also for $\beta\rightarrow\infty$.
These observations have important consequences in the parametrization of the weak capture matrix element $\Lambda$.

\end{document}